\documentclass[journal=jacsat,manuscript=article,suppfinfo]{achemso}
\pdfoutput=1

\usepackage[version=3]{mhchem} % Formula subscripts using \ce{}
\usepackage{lmodern}% allow to use arbitrary font size
\usepackage{graphicx} % constrain the graphic size
\usepackage{nicefrac}
\usepackage{microtype}
\usepackage[symbol*]{footmisc}
\usepackage{booktabs}
\usepackage{multirow}

\newcommand{\beginsupplement}{%
        \setcounter{table}{0}
        \renewcommand{\thetable}{S\arabic{table}}%
        \setcounter{figure}{0}
        \renewcommand{\thefigure}{S\arabic{figure}}
        \setcounter{equation}{0}
        \renewcommand{\theequation}{S\arabic{equation}}%
     }

\author{Yujin Tong}
\email{tong@fhi-berlin.mpg.de}
\phone{+49 (0)30 84135220}
\fax{+49 (0)30 84135206}
\author{Igor Ying Zhang}
\author{R. Kramer Campen}
\affiliation[FHI]
{Fritz Haber Institute of the Max Planck Society, 14195 Berlin, Germany}
\title[\ce{ClO4-}]
  {Experimentally Quantifying Anion Polarizability at the Air/Water Interface}


\begin{document}

%%%%%%%%%%%%%%%%%%%%%%%%%%%%%%%%%%%%%%%%%%%%%%%%%%%%%%%%%%%%%%%%%%%%%
%% The abstract environment will automatically gobble the contents
%% if an abstract is not used by the target journal.
%%%%%%%%%%%%%%%%%%%%%%%%%%%%%%%%%%%%%%%%%%%%%%%%%%%%%%%%%%%%%%%%%%%%%
\begin{abstract}
The adsorption of anions from aqueous solution on the air/water interface controls important heterogeneous chemistry in the atmosphere and is thought to have similar physics to anion adsorption at hydrophobic interfaces more generally. Starting in the mid 1990s a wide variety of theoretical and experimental approaches have found the adsorption of large, polarizable anions is thermodynamically favorable. While the qualitative insight is clear, determining the role of polarizability in adsorption has proven surprisingly challenging: simple physical models make clear that nonpolarizable anions will not adsorb, but trends in anion adsorption are difficult to rationalize based on polarizability and, in some theoretical approaches, adsorption is observed without change in anion polarization. Because there are no \emph{experimental} studies of interfacial anion polarizability, one possible explanation for this apparent contradiction is that theoretical descriptions suffer from systematic error. In this study we use interface specific nonlinear optical spectroscopy to extract the spectral response of the interfacial \ce{ClO4-} anion. We find that (i) the interfacial environment induces a break in symmetry of the anion due to its solvation anisotropy (ii) the Raman depolarization ratio, a measure of the ratio of the change in two components of \ce{ClO4-}'s polarizability tensor with change in nuclear positions, is $>2\times$ larger than its value in the adjoining bulk phase and interfacial concentration dependent and (iii) using a simple theoretical description we find that our measured changes in depolarization ratio are consistent with known changes in surface potential and tension with small changes in bulk \ce{ClO4-} concentration. The notion that interfacial anion polarizability differs from that in bulk, and that this polarizability is coverage dependent, is not accounted for in any current theoretical treatment of ions at the air/water interface. Accounting for such effects in classical models or validating their reproduction in \textit{ab initio} would be a valuable next step in understanding the physics of anion adsorption at the air/water interface.

\end{abstract}

%%%%%%%%%%%%%%%%%%%%%%%%%%%%%%%%%%%%%%%%%%%%%%%%%%%%%%%%%%%%%%%%%%%%%
%% Start the main part of the manuscript here.
%%%%%%%%%%%%%%%%%%%%%%%%%%%%%%%%%%%%%%%%%%%%%%%%%%%%%%%%%%%%%%%%%%%%%
\section{Introduction}

The adsorption of anions on hydrophobic interfaces controls important 
chemistry on aerosol surfaces and determines the stability of proteins, 
colloids and foams in a wide variety of environmental, physiological
and engineered settings. Anion adsorption on the air/water interface, the paradigmatic hydrophobic surface, has been particularly
well studied \cite{wei96, Ghosal05, Colussi06, Douglas06, Richard06,
Yukio08, Ghosal09, Allen13, Christopher13, Motschmann14, Allen15}. 
Perhaps the simplest question one can ask of this system is do anions
tend to exist at higher or lower concentrations at the air/water
interface than in the adjoining bulk aqueous phase: is the adsorption
of anions at the air/water interface thermodynamically favorable?

In principle measurements of surface tension of the air/water interface
as a function of bulk ionic strength provide such insight. Many
decades of such measurements have confirmed that surface tensions of
aqueous salt solutions increase with increasing ionic strength, those
of acids decrease, and that the magnitude of the effect depends strongly
on anion and only weakly on cation \cite{hey10,Randles57,ran66,wei96,
mat99,mat01}. Historically the first two observations were rationalized
by Wagner, Onsager and Samras (WOS) in their extension of the
Debye-H\"uckel theory of bulk aqueous electrolytes to interfaces. Within
this description anions tend to be \emph{excluded} from the air/water
interface because exposure to a low dielectric phase leads to an enormous,
unfavorable increase in electrostatic self-energy \cite{Wagner24, Samaras34}.
While qualitatively consistent with surface tension measurements, this
approach does not explain the strong dependence of the size of the measured
effect on anion type, nor the weak dependence on cation.   

Spurred by measured reaction rates of gas phase species with solvated halogen
anions in atmospheric aerosols that were too fast to be explained unless anions
exist at aersol surfaces in \emph{higher} concentration than in the adjoining 
bulk phase \cite{Knipping301}, subsequent work -- simulation using classical
polarizable force fields, various surface sensitive spectroscopies \cite{Ghosal05,
Padm07, Otten12, Allen15}, dielectric continuum theory \cite{Barry01, Levin09, 
Levin009, Wang14}, and properly parameterized fixed charge models \cite{net12}
 -- has shown this Debye-H\"uckel inspired view to be incorrect. Large polarizable
anions tend to exist in higher concentrations at the air/water interface than in
the adjoining bulk liquid. While the qualitative picture is clear, understanding
of quantitative trends in anion adsorption, \textit{e.g.}\ why does \ce{I-} adsorb
more strongly than \ce{Cl-}, and gaining atomically resolved insight into the
driving force of anion adsorption has proven extremely challenging.

Central to this challenge is resolving the contribution of ion polarizability to
the free energy of ion adsorption. Within the context of WOS theory ions are
represented as nonpolarizable point charges. As illustrated by Levin for an
idealized anion, it is this lack of polarizability that leads to the large
penalty in electrostatic self-energy for anion adsorption in the WOS model: for
an ideal, polarizable anion essentially all charge density shifts towards the
aqueous phase as the ion approaches the interface \cite{Levin09}. While it is
thus clear that polarizability must play a role in the thermodynamic driving
force of ion adsorption at hydrophobic interfaces, such an idealized model doesn't
offer a molecularly resolved or quantitative view of \emph{how}. Initial attempts
to describe ion adsorption in classical simulations concluded that explicit
description of ion polarizability was critical and that the relative surface
propensity of different ions was proportional to their polarizability and radius 
(\textit{i.e.}\ large, soft, polarizable anions more favorably adsorb) \cite{Douglas06,
Christopher13}. However, a variety of subsequent simulation studies have found
that anion adsorption occurs in properly parameterized classical models without
explicit description of polarizability, that relative anion polarization (where ion polarization is the product of applied field and polarizability
$p = \alpha\cdot E$) does not correlate with both experimental and simulated trends
in ion adsorption propensity, and that simulated interface active anions may be
similarly polarized in bulk water and at the interface \cite{net12,Caleman11,
Otten12, Baer11, Baer12, Sun15}. One possible explanation for these conflicting
results may lie in the difficulty in simulating the interfacial potential of the
(pure) water/air interface \cite{Baer12} (and thus error compensation between ion
polarizability and surface potential). A second is that, in general, theoretical
approaches describe anion polarizability at interfaces incorrectly because of a
lack of suitable experimental data against which classical polarizability models
might be parameterized or calculated \textit{ab initio} polarizabilities
validated. 

This missing experimental data is particularly important because one might expect anion
polarizability to change at the air/water interface. For monoatomic species this 
occurs because electron densities of ions at air/water interfaces must reflect 
the underlying asymmetry of electron density in the solvent, while for 
multiatomic anions additionally interface induced changes in nuclear arrangement 
(\textit{e.g.}\ bond lengths and angles) might be expected to enhance such 
effects. Furthermore, because such anion structural change might be expected to 
change as a function of interfacial potential, and because surface potential of 
the air/water interface is a function of the ionic strength of the bulk aqueous 
phase \cite{ran66,yan01}, it is possible to imagine that anion polarizability at 
interfaces may be a \emph{function} of bulk ionic strength.  
 
To find an experimental observable of \emph{interfacial} anion polarizability, 
it is useful to first consider how one might characterize ion polarizability in 
bulk liquid \ce{H2O}. The Raman depolarization ratio ($\rho$) is one such useful 
experimental constraint. Given an isotropic distribution of ions in liquid water 
and a molecular coordinate system (\textit{a,b,c}) in which \textit{a} (or 
\textit{b}) is taken to be perpendicular to the net deformation of a particular 
normal mode and \textit{c} parallel, one writes  \cite{hir92}:
\begin{equation}\label{e:Raman}
\rho = \frac{\text{I}_{\perp}}{\text{I}_{\parallel}} = \frac{3}{4 + 
		5\left(\frac{1+2R}{R-1}\right)^{2}}
\end{equation}  
where $\text{I}_{\perp}$ and $\text{I}_{\parallel}$ are the intensity of 
inelastic scattered light measured perpendicular and parallel to the, plane 
polarized incident field and $R=\frac{\nicefrac{\partial\alpha_{aa}^{(1)}}
{\partial Q}}{\nicefrac{\partial\alpha_{cc}^{(1)}}{\partial Q}}$. That is the 
Raman response of a particular mode can be quantitatively related to the change in 
the symmetry of the molecules polarizability tensor as the molecule is deformed 
in the mode's characteristic manner. Given this definition of $\rho$ is it perhaps 
unsurprising that several studies have shown that the ability to calculate the Raman response of mode that are strongly coupled to the environment is a sensitive test of the accuracy of the polarizability model employed \cite{ham14c,ito16}. 
Because spontaneous Raman is not interface-specific it is generally not possible 
to extract the $\rho$ of \emph{interfacial} anions. Clearly if we could, however, 
this observable could provide the sort of experimental constraint we seek.

Vibrationally resonant Sum Frequency (VSF) spectroscopy is a nonlinear optical, 
laser-based technique in which pulsed infrared and visible lasers are spatially 
and temporally overlapped at an interface and the output at the sum of the 
frequencies of the two incident beams monitored. The emitted VSF field is 
interface specific by its symmetry selection rules and a spectroscopy because as 
one tunes the frequency of one of the incident fields (in this case the infrared (ir)) in 
resonance with an optically accessible transition the intensity of the emitted 
sum frequency field ($\text{I}_{\text{sf}}$) increases by several orders of 
magnitude. Much prior work has shown that the intensity of the measured sum 
frequency response at a frequency $\omega$ is proportional to the change in 
polarizability ($\alpha_{ab}$) and dipole ($\mu_c$) with motion along the normal 
mode of frequency $\omega$:\cite{lam05}
\begin{equation}\label{e:SFG}
 \text{I}_{\text{sf}} \propto \chi^{(2)}_{ijk} \propto \beta^{(2)}_{abc} \propto - \frac{1}{2\epsilon_{0}\omega}\frac{\partial\alpha^{(1)}_{ab}}{\partial Q}\frac{\partial \mu_{c}}{\partial Q}
\end{equation} 
in which $\chi^{(2)}_{ijk}$ is the macroscopic nonlinear susceptibility in the 
lab coordinate system $(ijk)$, $\beta^{(2)}_{abc}$ the molecular 
hyperpolarizability and both are second rank tensors. Because by varying 
experimental conditions, \textit{i.e.}\ beam incident angles and field 
polarizations, one can selectively probe different components of $\beta^{(2)}$, 
a correctly chosen ratio of intensities allows the direct measurement of 
$R=\frac{\nicefrac{\partial\alpha_{aa}^{(1)}}{\partial Q}}{\nicefrac{\partial\alpha_{cc}^{(1)}}{\partial Q}}$, and thus the possibility of extracting the Raman depolarization ratio of anions with interfacial specificity. That is, by comparing measurements of $\rho$ for an anion in solution and at the air/water interface experimental estimates of anion polarizability anisotropy at aqueous interfaces (and the change in anion polarizability anisotropy on moving from bulk liquid water to the aqueous interface) are possible.

The perchlorate anion (\ce{ClO4-}) has tetrahedral ($T_{d}$) symmetry in bulk 
liquid \ce{H2O} and a favorable free energy of adsorption at the air/water 
interface \cite{Irish1984,Christopher13}. Here, using VSF spectroscopy, we probe 
two Cl-O modes of the perchlorate anion at the air/water interface and show (i) 
that the T$_{d}$ symmetry of the \ce{ClO4-} anion in bulk water is lifted at this 
interface: the \ce{ClO4-}'s $\nu_{1}$ mode, that is IR inactive in bulk, is now 
active, and (ii) that the Raman depolarization ratio of the $\nu_{1}$ mode 
increases monotonically, and by more than $2\times$, with increasing bulk 
concentration. Put another way, the polarizability tensor of interfacial 
\ce{ClO4-} grows increasingly anisotropic with increasing interfacial population. 
Using a simple computational model we show that the increase in polarizability 
anisotropy with increasing bulk concentration we observe in experiment can be 
quantitatively related to increases in interfacial field (consistent with prior 
measurements of concentration dependent surface potential \cite{mar16c}), 
\ce{ClO4-} dipole, and relative bond length of one Cl-O bond with respect to the 
other three with increasing concentrations of bulk \ce{HClO4}. Quantitative 
theoretical insights into the driving force of anion adsorption at the air/water 
interface, and specific ion effects more generally, require accurate calculation of ion polarizability at aqueous interfaces. The results of this study are the first, of 
hopefully many, experimental observations of this ionic property.

\section{Results and discussion}
\begin{figure}[ht]
\includegraphics[width=16cm,height=16cm,keepaspectratio]{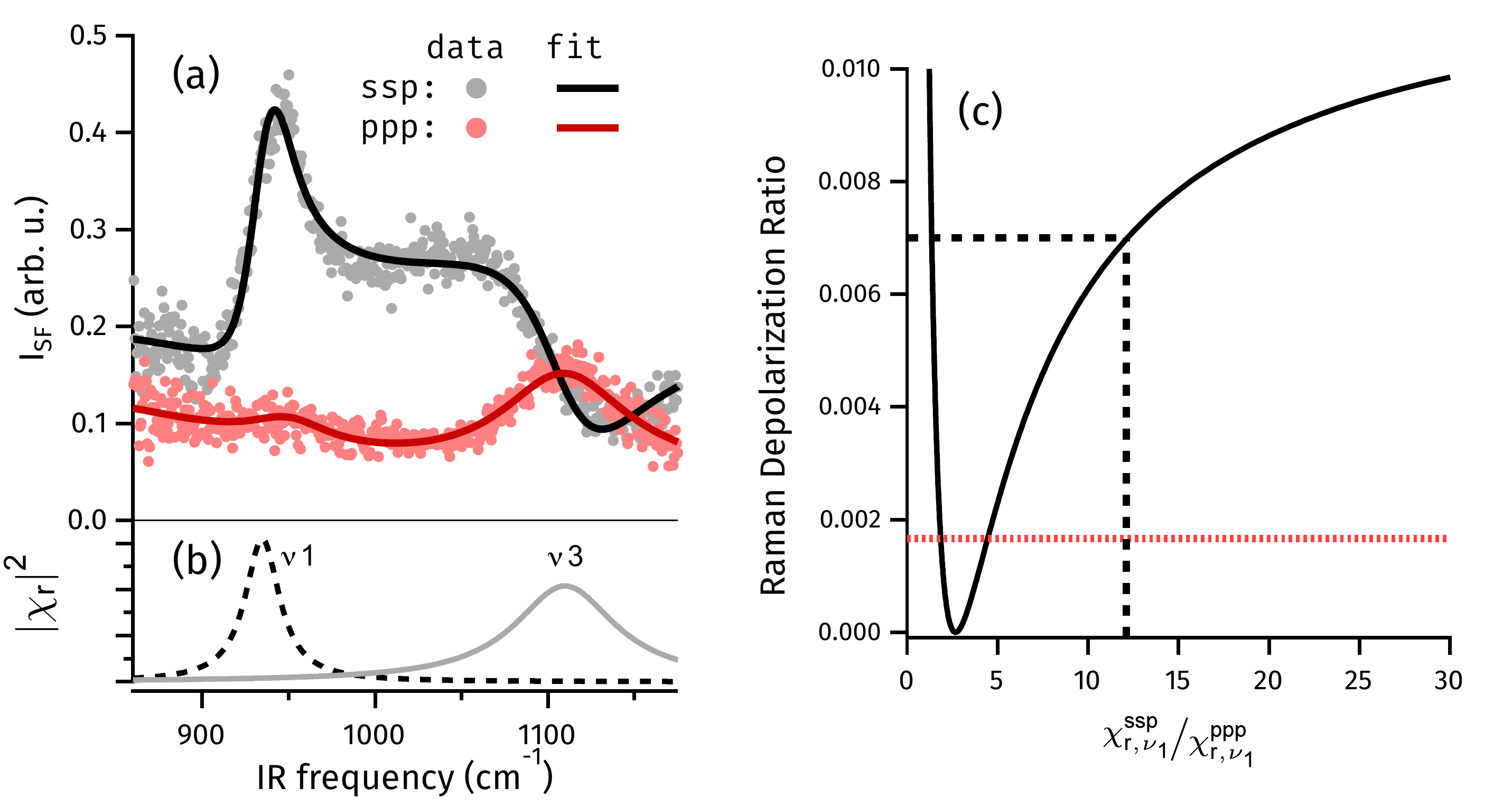}
\caption{a.) SFG spectrum of 0.6 M \ce{HClO4} solution at the air/water
 interface measured under \textit{ssp} (black traces) and \textit{ppp}
(red traces) polarization combinations. Circular symbols are the experimental
observations; solid traces are the fit to equation \ref{e:line}; b.) Two components obtained
from spectral fit, assigned to $\nu_{1}$ (A1 symmetry) and $\nu_{3}$ (F2 symmetry)
vibrational modes of perchlorate respectively; c.) Calculated Raman
depolarization ratio (black solid line) as a function of the measured VSF peak
amplitude ratio ($\nicefrac{\chi^{\mathsf{ssp}}_{\mathsf{r,}\nu_{1}}}{\chi^{\mathsf{ppp}}_{\mathsf{r,}\nu_{1}}}$); the black dashed line is the result of fitting the data shown in (a), the red dashed line indicates the bulk value reported by prior authors. \cite{Hyodo1989}}
\label{fig1}
\end{figure}

Figure \ref{fig1}(a) shows the VSF spectra from the air/0.6 M \ce{HClO4} solution 
interface measured under the \textit{ssp} (\textit{s}-polarized SF, 
\textit{s}-polarized visible, and \textit{p}-polarized IR ) (black circles) and 
\textit{ppp} (red squares) polarization combinations. There are two resonances 
apparent in this frequency range. Fitting both spectra simultaneously with the 
Lorentzian lineshape model described in the methods section results in resonances 
centered at 935 and 1110 cm$^{-1}$ (Figure \ref{fig1}(b) dotted lines) 
\footnote{It is clear from inspection that the apparent baseline of the spectrum 
is higher on the low than high frequency side of the data shown in Figure 
\ref{fig1}. We have previously showed that the libration of water at the 
air/water interface is a broad spectral feature centered at ~834 cm$^{-1}$ 
(\textit{i.e.}\ well outside the spectral window of the current study) 
\cite{tong16PCCP}. In performing the global fit of the data in Figure \ref{fig1} 
we included the libration as described in the SI.}. 

Because both spectral features are absent in pure water, and are spectrally 
separated from resonances of water or likely impurities, they can be 
straightforwardly assigned by reference to Raman and IR measurements of bulk 
aqueous perchloric acid and perchlorate salt solutions \cite{Irish1984, Hyodo1989,
Grigorovich1976}. In brief, the \ce{ClO4-} anion has four normal modes apparent 
in calculation: the $\nu_{1}$ at 930, the $\nu_{2}$ at 450, the $\nu_{3}$ at 1100 
and the $\nu_{4}$ at 620 cm$^{-1}$. All four are Raman active at all 
concentrations in bulk aqueous solution but the $\nu_{1}$ and $\nu_{2}$ are only 
apparent in IR absorption spectra at bulk concentrations greater than $\approx$ 
11 M. This observation is a straightforward consequence of anion symmetry: at 
bulk concentrations below 11 M the \ce{ClO4-} anion has T$_{d}$ symmetry (under 
which condition $\nu_{1}$ and $\nu_{2}$ are IR inactive) and at sufficiently high 
concentrations this symmetry is broken: either by ion pairing or, in the case of 
\ce{HClO4}, by the presence of molecular acid. If we assign the resonance 
apparent in Figure \ref{fig1}(b) at 935 cm$^{-1}$ to the $\nu_{1}$ mode and that 
apparent at 1110 cm$^{-1}$ to the $\nu_{3}$, we are left with an apparent 
incongruity. As shown in equation \ref{e:SFG}, VSF activity requires that a mode 
must be \emph{both} IR \emph{and} Raman active. This implies the \ce{ClO4-} anion 
must lose its T$_{d}$ symmetry at the air/water interface at concentrations that 
are at least 15$\times$ lower than those at which T$_{d}$ symmetry is lifted in 
bulk. 

We imagine three possible mechanisms for the loss of T$_{d}$ symmetry: consistent 
with recent work on other strong acids, molecular \ce{HClO4} may exist at the 
air/water interface at concentrations dramatically lower than in bulk 
\cite{Baer12a}, \ce{ClO4-} may no longer have tetrahedral symmetry due to ion 
pairing, or it may not have tetrahedral symmetry due to, more general, solvation anisotropy at the interface. We tested the first possibility by collecting spectra from 0.6 M solutions of \ce{NaClO4}. For this a similarly intense $\nu_{1}$ feature is observed suggesting the likely cause of T$_{d}$ symmetry lifting is not molecular acid (see Electronic Supplementary Information for data).

To evaluate the possibility of T$_{d}$ symmetry lifting due to ion pairing, 
it is necessary to understand how ion pairing might be expected to influence the 
$\nu_{1}$ and $\nu_{3}$ spectral response. In bulk solutions of perchlorate salts 
at concentrations above 1 M the center frequency of perchlorate's $\nu_{3}$ mode 
has been observed to continuously shift as a function of concentration 
\cite{che04b}. This concentration dependent spectral evolution has been assigned 
to the formation of weak, solvent-separated ion pairs. As mentioned above, in 
this concentration range the $\nu_{1}$ mode is infrared inactive. At 
still higher concentrations in bulk water, $>$ 11 M, perchlorate's degenerate 
modes, \textit{i.e.}\ $\nu_{2}$, $\nu_{3}$ and $\nu_{4}$, have been observed to 
split due to contact ion pair formation, where the degree of splitting is a 
function of the extent to which symmetry is broken \cite{rit75,che04b}. 

\begin{figure}[ht]
\includegraphics[width=0.8\textwidth,keepaspectratio]{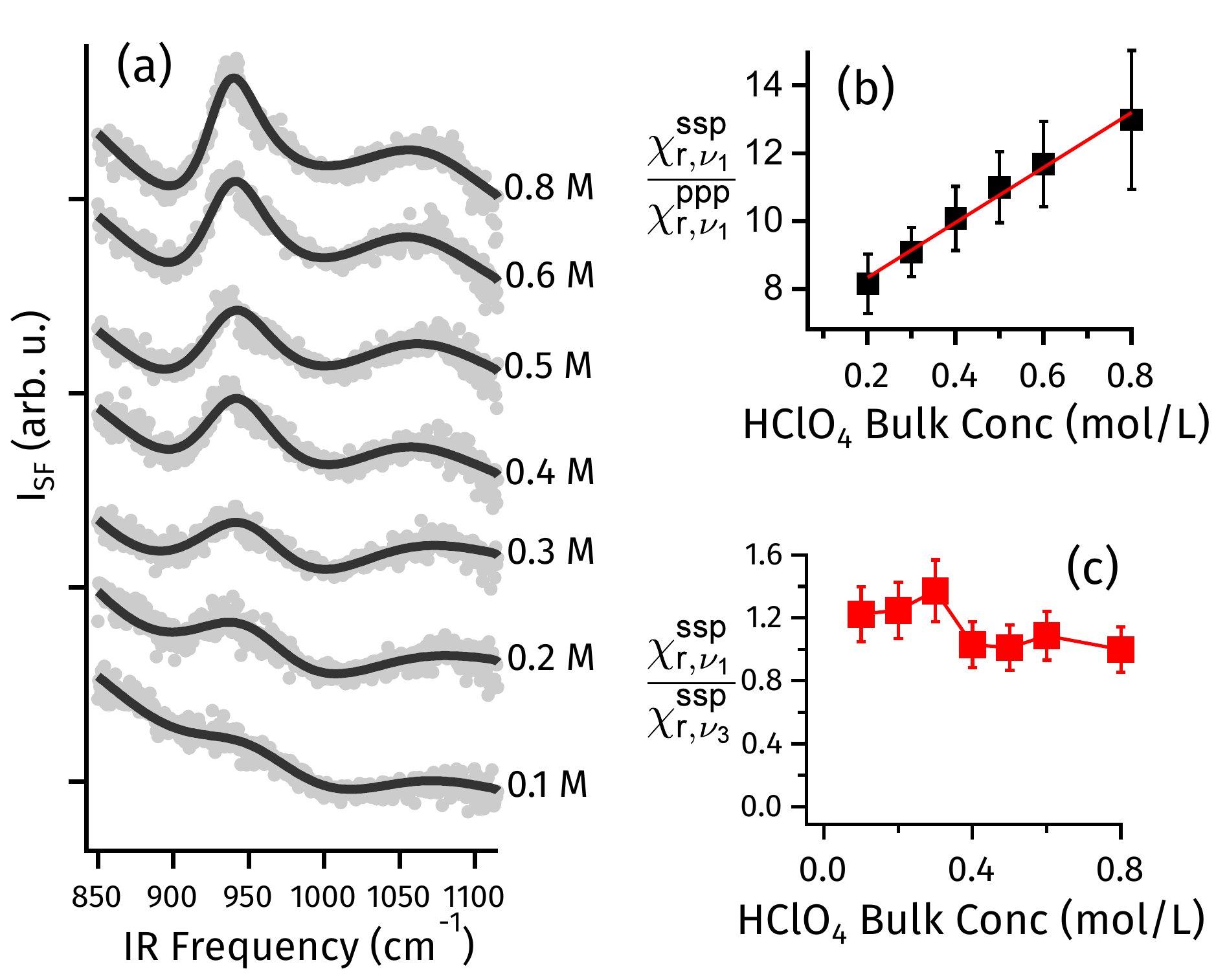}
\caption{a) VSF spectra as a function of \ce{ClO4-} concentration below 1 M for 
the \textit{ssp} polarization combinations. Grey dots are data and solid black 
lines the corresponding fits. b) $\nicefrac{\chi^{\mathsf{ssp}}_{\mathsf{r},
\nu_{1}}}{\chi^{\mathsf{ppp}}_{\mathsf{r},\nu_{1}}}$ for the data in a) and the SI 
illustrating the linearity of this ratio with respect to bulk \ce{HClO4} 
concentration. The uncertainty at each point is extracted from the fits to the 
data as described in the SI. 
c) $\nicefrac{\chi^{\mathsf{ssp}}_{\mathsf{r},\nu_{1}}}{\chi^{\mathsf{ssp}}
_{\mathsf{r},\nu_{3}}}$ extracted from the data in a) and the SI. Clearly this 
ratio is essentially constant with respect to bulk concentration of \ce{HClO4}. As 
discussed in the text this indicates that the orientation of interfacial 
\ce{ClO4-} is concentration independent. The straight line is a guide to the
eye.}
\label{fig2}
\end{figure}

As is discussed in detail below (see Figure \ref{fig2} for data) at bulk 
concentrations lower than 1 M \ce{HClO4} the $\nu_{1}$ mode is clearly VSF (and 
thus IR) active, the $\nu_{3}$ spectral response is quantitatively reproduced 
with a single center frequency and line width: splitting or frequency shift of 
the $\nu_{3}$ resonance is not required to describe our data. We therefore 
conclude that neither contact nor solvent separated ion pair formation 
explains the lifting of T$_{d}$ symmetry. Given that VSF spectra collected at 
bulk concentrations below 1 M \ce{HClO4} are consistent with formation of neither 
weak, solvent separated ion pairs nor contact ion pairs, we conclude that the 
lifting of T$_{d}$ symmetry for interfacial \ce{ClO4-} (and thus the IR and VSF 
activity of the $\nu_{1}$ mode) must be the result of the intrinsic anisotropy of 
the solvation environment at the air/water interface: solvation anisotropy must 
induce sufficient structural deformation in the \ce{ClO4-} anion to lift the bulk 
T$_{d}$ symmetry and make the $\nu_{1}$ mode IR, and VSF, active.

While this qualitative observation of a consequence of structural deformation is 
important, to make clear connections to theory it would be useful to quantify 
this deformation, and the resulting change in the perchlorate polarizability 
tensor. As shown in the Electronic Supplementary Information, given the $\nu_{1}$ spectral 
amplitudes extracted from the fit of the \textit{ssp} and \textit{ppp} data in 
Figure \ref{fig1}(a), and assuming the \ce{ClO4-} anion is oriented such that one 
Cl-O points along the surface normal and that the anion has C$_{3\nu}$ symmetry, 
we can calculate the Raman depolarization ratio, \textit{i.e.}\ $\rho$, for 
perchlorate ions at the interface. The result of this calculation is shown by the 
solid black line in Figure \ref{fig1}(b). The ratio of spectral amplitudes of the 
$\nu_{1}$ \textit{ssp} and \textit{ppp} spectral amplitudes shown in Figure 
\ref{fig1}(a) suggests a $\nu_{1}$ $\rho$ of $0.0063$ or more than $2\times$ 
larger than the same quantity for \ce{ClO4-} in bulk. 

It is worth emphasizing that the result that interfacial \ce{ClO4-} is significantly larger than bulk is insensitive to the simplifying assumptions required in its calculation. As we show in the Electronic Supplementary Information, assuming the \ce{ClO4-} is oriented with one Cl-O bond at an increasing, nonzero angle with respect to the surface normal leads to slightly \textit{larger} estimates for the depolarization ratio of $\nu_{1}$ while assuming the symmetry of the \ce{ClO4-} anion has decreased to C$_{2\nu}$ or C$_{\infty\nu}$ leads to quantitatively similar results. Note also that because we measure the \emph{intensity} of the emitted sum frequency light, and not the \emph{field}, our measurements are equally consistent with one Cl-O bond along the surface normal and the remaining three oxygens pointing either towards the bulk liquid or air. Prior theoretical studies imply, at least in the low concentration limit, that the latter configuration is favored \cite{Ghosal09,ott09,hey16}.   

As noted above, changes in polarizability must be correlated with changes in anion 
nuclear structure. As we show in detail in the Electronic Supplementary Information, a simple 
computational model suggests a $\rho$ of 0.0062 is consistent with a 3\% change in 
Cl-O bond length (for the Cl-O along the surface normal), and a permanent dipole 
moment of interfacial \ce{ClO4-} of 0.75 Debye (\textit{n.b.}\ consistent with the 
absence of IR active $\nu_{1}$ and $\nu_{2}$ modes for the \ce{ClO4-} anion in 
vacuum or in bulk liquid water perchlorate's dipole moment in either bulk phase is 
below our detection limit). In this manner our experimental observable directly 
constrains the anisotropy of interfacial perchlorate's polarizability tensor and 
places quantitative constraints on interfacial \ce{ClO4-} polarization at the air/
water interface.

To gain more insight into the fate of the \ce{ClO4-} anion at the air/water 
interface we next measured its $\nu_{1}$ and $\nu_{3}$ spectral amplitudes as a 
function of bulk concentration of \ce{HClO4} from 0.1 - 0.8 M (higher 
concentrations lead to qualitative change in the spectral response, possibly the 
result of interface induced ion pairing, as shown in the Electronic Supplementary Information). 
The concentration dependent spectra collected under the \textit{ssp} polarization condition 
are shown in Figure \ref{fig2}(a), the concentration dependent \textit{ppp} are 
plotted in the Electronic Supplementary Information. Global fitting of both sets of data allows 
the extraction of the $\nicefrac{\chi_{\mathsf{r,}\nu_{1}}^{\mathsf{ssp}}}{\chi_{\mathsf{r,}\nu_{1}}^{\mathsf{ppp}}}$ ratio 
as a function of concentration. As shown in Figure \ref{fig2}(b) changing bulk 
concentrations of \ce{HClO4} from 0.1 - 0.8 M leads to a change of this ratio from 
8 - 13. Reference to Figure \ref{fig1} makes clear that this change in spectral 
amplitude implies an increase in the raman depolarization ratio of $\nu_{1}$ of 
0.004 - 0.007 over the same concentration range. Evidently, with increasing 
interfacial population, \ce{ClO4-} polarizability grows increasingly anisotropic. 
Using the same simple computational model discussed above, increasing the 
depolarization ratio from 0.004 to 0.007 is consistent with an increase in 
interfacial field from 139--171 (meV), an elongation in the Cl-O bond along the 
surface normal of 2.6--3.3 \% and a change in \ce{ClO4-} dipole moment from 
0.6--0.76.

The relationship between $\nicefrac{\chi_{\mathsf{r,}\nu_{1}}^{\mathsf{ssp}}}{\chi_{\mathsf{r,}\nu_{1}}^{\mathsf{ppp}}}$ and 
Raman depolarization ratio shown in Figure \ref{fig1} assumes that the \ce{ClO4-} 
is orientated such that one Cl-O group points along the surface normal. Applying 
this analysis to the data shown in Figure \ref{fig2}(a) implicitly assumes that 
this orientation is concentration independent. Because the $\nu_{1}$ and $\nu_{3}$ 
normal modes are orthogonal, we would expect any concentration dependent change in 
the orientation of interfacial \ce{ClO4-} to result in significant change in the 
$\nicefrac{\chi_{\mathsf{r,}\nu_{1}}^{\mathsf{ssp}}}{\chi_{\mathsf{r,}\nu_{3}}^{\mathsf{ssp}}}$ (see Electronic Supplementary Information for a calculation of the size of this effect). As is shown in Figure 
\ref{fig2}(c) this is not the case. We thus conclude that the orientation of 
interfacial \ce{ClO4-} is, to within the limits of our sensitivity, over 0.1-0.8 M 
range in bulk concentration, concentration independent.

Our results suggest the following model for \ce{ClO4-} at the air/water interface: 
on adsorption \ce{ClO4-} is polarized, \textit{i.e.}\ it has a nonzero dipole 
moment, and the polarizability anisotropy changes due to a change in the bond 
length of the Cl-O that points along the surface normal relative to the three 
other Cl-O bonds. With increasing interfacial concentrations of \ce{ClO4-} the 
interfacial field increases, ion polarization increases (the dipole continues to 
grow) and the polarizability anisotropy continues to increase.

Modern dielectric continuum descriptions, principally developed over the last 
eight years by Levin and coworkers, largely reproduce experimentally measured 
changes in surface tension with increasing concentrations of both acids and salts 
\cite{Levin009}. In this approach anions are treated as spheres whose 
polarizability and radius are concentration dependent input parameters and whose 
electrostatic self energy is defined relative to the dielectric constants of, 
bulk, water and air. Notably the largest disagreements between experiment and 
theory exist for \ce{ClO4-} solutions (both acids and salts). Levin, dos Santos 
and coworkers have suggested that this is likely the result of inaccuracies in the 
estimates of ionic radii for \ce{ClO4-} \cite{Santos10,lev14}. Our results are 
consistent with an alternative scenario in which anion polarizability (and 
\ce{ClO4-} radius) is interfacial concentration dependent. While our results imply 
the relationship between \ce{ClO4-} radius and interfacial concentration is 
monotonic, larger multivalent ions might be expected to have a more complicated 
interplay between polarizability, structure, dipole and interfacial concentration.

Atomistic simulation studies, whether employing classical or ab-initio potential 
energy surfaces, have largely reported either potentials of mean force for ion 
adsorption in the limit of infinite dilution or \textit{brute force} simulations 
at a fixed ion concentration. As alluded to above, while important and informative 
these studies suffer from a variety several possible shortcomings. The lack of 
experimental constraints on polarizability means that there is no experimental 
parameterization of classical polarizability models and that \textit{ab-initio} 
treatments of polarizability cannot be validated. To further heighten the 
challenge the surface potential of pure water is both difficult to measure 
experimentally and the subject of significant disagreement, (by more than 0.5 V) 
in simulation treatments \cite{kat11}. Thus one might expect that inaccuracies in surface 
potential of the pure water/air interface might, plausibly compensate for 
inaccuracies in polarizability treatment.

Data of the sort described in this study gives a clear path forward through these 
challenges. Clearly, given experimental constraints on interfacial anion 
polarizability, empirical polarizability models can be more accurately 
parameterized and \textit{ab-initio} treatments validated. Given this validated 
polarizability model, systematically reducing errors in the calculation of (ion concentration dependent) surface potential is now much more straight forward.

Prior workers have performed studies similar in spirit to those shown here. 
Miyame, Morita and Ouchi characterized the S-O stretch vibrations of SO$_{4}^{2-}
$, while Motschmann and coworkers characterized the CN stretch vibrations of the 
potassium ferricynaide ion, \textit{i.e.}\ Fe(CN)$_{6}^{4-}$, as a function of bulk 
concentration at the air/water interface \cite{Yukio08,Motschmann14}. Consistent 
with both calculation and other experimental approaches that suggest SO$_{4}^{2-}$ retains its bulk solvation shell at the air/water interface \cite{gop06,Douglas06}, Miyame, Morita and Ouchi 
find interfacial SO$_{4}^{2-}$ to be essentially the same as bulk. In contrast 
Motschmann and coworkers found that CN modes that were IR inactive in bulk 
solution were apparent in the VSF spectrum, \textit{i.e.}\ the interface induces a 
change in ferricyanide symmetry as it does for perchlorate. However, presumably 
because of the more structurally complicated anion, they were unable to quantify 
the resulting change in the polarizability tensor.

As is clear from equation \ref{e:SFG} if a molecules hyperpolarizability is  
concentration independent it should, in principle, be possible to extract a 
measurement of anion interfacial density as a function of bulk concentration by 
plotting the square of the measured SF signal vs. bulk concentration. In a series 
of studies employing electronically resonant second harmonic measurements of a 
variety of anions at air/water interface Saykally and co-workers have treated 
anion hyperpolarizability as concentration independent, fit adsorption isotherms 
to measurements of SHG signal as a function of bulk concentration, and calculated 
anion adsorption energies \cite{Richard06,Otten12}. Our results suggest that this 
type of data needs to be revisited. Because deformation of the perchlorate leads 
to an \emph{increase} in dipole and polarizability tensor ratio it is clear that, 
given a VSF spectra collected under the \textit{ssp} polarization condition, using 
this approach would significantly overestimate adsorption energies (\textit{i.e.}\ 
the molecular response of the perchlorate anion would increase with increasing 
concentration). 

In summary, in the current study we have employed VSF spectroscopy and a simple 
computational model to study the behavior of \ce{ClO4-} at the air/\ce{HClO4} 
solution interface. Consistent with much prior work our observations clearly 
demonstrate that \ce{ClO4-} is a surface active anion. We significantly extend 
these prior efforts by demonstrating that the presence of the interface induces 
deformation of the anion, which cause a bulk forbidden mode to be VSF active due 
to change in anion symmetry, creates a nonzero dipole moment and leads to a change 
in the measured polarizability anisotropy. Our results suggest that increasing 
density of \ce{ClO4-} at the interface leads an increasing interfacial field that 
both leads to increasing \ce{ClO4-} polarization (\textit{i.e.}\ increasing 
\ce{ClO4-} dipole moment), and increasingly anisotropic \ce{ClO4-} polarizability 
\cite{Douglas06, Richard06, Christopher13}. Extension of the approach we describe 
here should allow the possibility of directly quantifying of the elements of 
\ce{ClO4-} polarizability tensor, rather than just their ratio, and allow straight 
forward estimate of any thermodynamic significance of the concentration dependent 
interfacial anion polarizability. 

The close connection we describe here between the dipole moment, structure and 
polarizability of interfacial anions with increasing interfacial field has not, to 
our knowledge, been previously considered but should be a quite general feature of 
anion, particularly polyvalent anion, adsorption at hydrophobic interfaces. As 
such its quantitative reproduction is a prerequisite for simulation approaches 
that attempt to offer microscopic insight into this phenomena.

\section{Methods}

\subsection{Solution Preparation}
\ce{HClO4} (Suprapur, 70\%, Merck) and \ce{NaClO4} (\textgreater 99.99\%,
Sigma-Aldrich) were used as received. Solutions with indicated concentration
were prepared by diluting the high concentration of \ce{HClO4} and \ce{NaClO4}
in ultrapure \ce{H2O} (18.3 M$\Omega\cdot$cm, Milli-Q, Millipore). All
solutions are prepared freshly before each measurement to limit degradation or
contamination. VSF measurements in the C-H stretching and C=O region were
employed to judge the quality of the solutions.

\subsection{VSF Measurement and Spectral Modeling}
The VSF spectrometer employed for the current measurement, and in particular its power at long infrared wavelengths has been described in detail in our previous
studies\cite{tong15, tong16PCCP}. In the interest of brevity only a brief description that pertinent to
this measurement will be given here. The IR beam was generated from a
commercial optical parametric amplifier (HE-TOPAS, Light Conversion) with a difference frequency generation (DFG) module. The full
width half maximum (FWHM) of the beam at frequency region between 600--1200
cm$^{-1}$ is typically 300 cm$^{-1}$ with GaSe was used as the DFG crystal. To
probe the interfacial Cl-O stretch modes the center frequency of the beam was
tuned to $\approx$ 1000 cm$^{-1}$.  A narrow-band visible (VIS) pulse was produced
from a home-made spectral shaper\cite{tong15}. The beam is centered at 800 nm
with a bandwidth of 15 cm$^{-1}$. The energy per pulse of the IR and VIS at the
sample surface was 5.8 and 15.4 $\mu$J respectively. Polarizations and energies
of the incident fields at the interface were controlled using $\lambda$/2 plate,
polarizer, $\lambda$/2 plate combinations. The two beams propagate in a
coplanar fashion and focused on the samples using lenses with focal lengths of
10 and 25 cm and incident angles of $39.5\pm 0.5^{\circ}$ and
$65\pm 0.5^{\circ}$ for the IR and VIS. All measurements were conducted in
ambient conditions at room temperature and under the \textit{ssp} (\textit{s}-polarized SF, \textit{s}-polarized visible, and \textit{p}-
polarized IR where \textit{p} indicates polarization in the plane of incidence
and \textit{s} polarization orthogonal) and \textit{ppp} polarization condition.
Non-resonant signals from a gold thin film were used to correct for the
frequency dependent IR intensity. The acquisition time for spectra of the gold
reference and samples were 30 and 300 s, respectively.

To quantify the observed VSF spectral response, we adopted a Lorentzian line shape model
described and justified in much previous work by us and others
\cite{tong13, tong15, tong16PCCP, bain91, zhuang99}. 
\begin{equation}\label{e:line}
	\text{I}_{\text{sf}}(\omega_{\text{sf}}) \propto \left|\chi^{(2)}_{\text{eff}} \right|^2
		\propto  \left|\left|\chi_{\text{nr}} \right|e^{i\epsilon} +
    \sum_{n}\frac{\chi_{\text{r},n}}{\omega_{\text{ir}} - \omega_{n}
    + i\Gamma_{n}}\right|^2
\end{equation}
where $\text{I}_{\text{sf}}(\omega_{\text{sf}})$ is the normalized VSF intensity,
$\chi^{(2)}_{\text{eff}}$ is the effective second order susceptibility, which
depends on the experimental geometry, molecular hyperpolarizability and
orientation.$\left|\chi_{\text{nr}}\right|$ and $\epsilon$ are the nonresonant
amplitude and phase and $\chi_{\text{r},n}$, $\omega_{n}$ and $\Gamma_{n}$ are the
complex amplitude, center frequency and line width of the \textit{n}$^{th}$ resonance.

To actually analyze the data we fit the measured VSF spectrum using the
Levenberg-Marquardt algorithm as implemented in the commercial visualization
and analysis program Igor Pro (Wavemetrics). Fitting spectra collected at each bulk concentration and polarization with this line shape model results in an underdetermined minimization problem. Because bulk studies suggest that the center frequencies and spectral shape of \ce{ClO4-} solution are concentration independent, we addressed this data by assuming that all spectra collected under a bulk concentration of \ce{HClO4} could be described with two resonances, each with  a concentration independent line width, center frequency and phase, and a nonresonant amplitude and phase that are also concentration independent. We accounted for the libration tail (only important in the \textit{ssp} spectra) by assuming the libration has the center frequency and line width from our previous study \cite{tong16PCCP}. Details of the analysis, and all the parameters resulting from the fit, are given in the Electronic Supplementary Information.

%%%%%%%%%%%%%%%%%%%%%%%%%%%%%%%%%%%%%%%%%%%%%%%%%%%%%%%%%%%%%%%%%%%%%
%% The "Acknowledgement" section can be given in all manuscript
%% classes.  This should be given within the "acknowledgement"
%% environment, which will make the correct section or running title.
%%%%%%%%%%%%%%%%%%%%%%%%%%%%%%%%%%%%%%%%%%%%%%%%%%%%%%%%%%%%%%%%%%%%%
\begin{acknowledgement}

The authors thank Martin Wolf for useful discussions and the Max Planck Society 
for support of this work.

\end{acknowledgement}

%%%%%%%%%%%%%%%%%%%%%%%%%%%%%%%%%%%%%%%%%%%%%%%%%%%%%%%%%%%%%%%%%%%%%
%% The same is true for Supporting Information, which should use the
%% suppinfo environment.
%%%%%%%%%%%%%%%%%%%%%%%%%%%%%%%%%%%%%%%%%%%%%%%%%%%%%%%%%%%%%%%%%%%%%
%\begin{suppinfo}
\subsection{Electronic Supplementary Information}

VSF spectra of 0.6 M perchlorate salt solution, details of line shape analysis 
and all data fitting results, full description of the theory describing the Raman 
depolarization ratio and its connection to VSF measurements, calculated dependence 
of the Raman depolarization ratio on \ce{ClO4-} orientation, calculated dependence 
of the $\nicefrac{\chi^{\mathsf{ssp}}_{\mathsf{r},\nu_{1}}}{\chi^{\mathsf{ssp}}
_{\mathsf{r},\nu_{3}}}$ on \ce{ClO4-} orientation, details of the electronic structure calculation and 
tables of all calculated results illustrating their independence of basis set.

%\end{suppinfo}

\subsection{Author Contributions}
YT and RKC designed the study. YT performed the measurements. YT and RKC analyzed the data and wrote the paper. IYZ performed the electronic structure calculations. All authors edited the paper. 

%%%%%%%%%%%%%%%%%%%%%%%%%%%%%%%%%%%%%%%%%%%%%%%%%%%%%%%%%%%%%%%%%%%%%
%% The appropriate \bibliography command should be placed here.
%% Notice that the class file automatically sets \bibliographystyle
%% and also names the section correctly.
%%%%%%%%%%%%%%%%%%%%%%%%%%%%%%%%%%%%%%%%%%%%%%%%%%%%%%%%%%%%%%%%%%%%%
%\bibliography{REF_ClO4}

\begin{mcitethebibliography}{54}
\providecommand*\natexlab[1]{#1}
\providecommand*\mciteSetBstSublistMode[1]{}
\providecommand*\mciteSetBstMaxWidthForm[2]{}
\providecommand*\mciteBstWouldAddEndPuncttrue
  {\def\EndOfBibitem{\unskip.}}
\providecommand*\mciteBstWouldAddEndPunctfalse
  {\let\EndOfBibitem\relax}
\providecommand*\mciteSetBstMidEndSepPunct[3]{}
\providecommand*\mciteSetBstSublistLabelBeginEnd[3]{}
\providecommand*\EndOfBibitem{}
\mciteSetBstSublistMode{f}
\mciteSetBstMaxWidthForm{subitem}{(\alph{mcitesubitemcount})}
\mciteSetBstSublistLabelBeginEnd
  {\mcitemaxwidthsubitemform\space}
  {\relax}
  {\relax}

\bibitem[Weissenborn and Pugh(1996)Weissenborn, and Pugh]{wei96}
Weissenborn,~P.~K.; Pugh,~R.~J. \emph{J Colloid Interface Sci} \textbf{1996},
  \emph{184}, 550--563\relax
\mciteBstWouldAddEndPuncttrue
\mciteSetBstMidEndSepPunct{\mcitedefaultmidpunct}
{\mcitedefaultendpunct}{\mcitedefaultseppunct}\relax
\EndOfBibitem
\bibitem[Ghosal \latin{et~al.}(2005)Ghosal, Hemminger, Bluhm, Mun, Hebenstreit,
  Ketteler, Ogletree, Requejo, and Salmeron]{Ghosal05}
Ghosal,~S.; Hemminger,~J.~C.; Bluhm,~H.; Mun,~B.~S.; Hebenstreit,~E. L.~D.;
  Ketteler,~G.; Ogletree,~D.~F.; Requejo,~F.~G.; Salmeron,~M. \emph{Science}
  \textbf{2005}, \emph{307}, 563--566\relax
\mciteBstWouldAddEndPuncttrue
\mciteSetBstMidEndSepPunct{\mcitedefaultmidpunct}
{\mcitedefaultendpunct}{\mcitedefaultseppunct}\relax
\EndOfBibitem
\bibitem[Cheng \latin{et~al.}(2006)Cheng, Vecitis, Hoffmann, and
  Colussi]{Colussi06}
Cheng,~J.; Vecitis,~C.~D.; Hoffmann,~M.; Colussi,~A. \emph{J Phys Chem B}
  \textbf{2006}, \emph{110}, 25598--25602\relax
\mciteBstWouldAddEndPuncttrue
\mciteSetBstMidEndSepPunct{\mcitedefaultmidpunct}
{\mcitedefaultendpunct}{\mcitedefaultseppunct}\relax
\EndOfBibitem
\bibitem[Jungwirth and Tobias(2006)Jungwirth, and Tobias]{Douglas06}
Jungwirth,~P.; Tobias,~D.~J. \emph{Chem Rev} \textbf{2006}, \emph{106},
  1259--1281\relax
\mciteBstWouldAddEndPuncttrue
\mciteSetBstMidEndSepPunct{\mcitedefaultmidpunct}
{\mcitedefaultendpunct}{\mcitedefaultseppunct}\relax
\EndOfBibitem
\bibitem[Petersen and Saykally(2006)Petersen, and Saykally]{Richard06}
Petersen,~P.~B.; Saykally,~R.~J. \emph{Ann Rev Phys Chem} \textbf{2006},
  \emph{57}, 333--364\relax
\mciteBstWouldAddEndPuncttrue
\mciteSetBstMidEndSepPunct{\mcitedefaultmidpunct}
{\mcitedefaultendpunct}{\mcitedefaultseppunct}\relax
\EndOfBibitem
\bibitem[Miyamae \latin{et~al.}(2008)Miyamae, Morita, and Ouchi]{Yukio08}
Miyamae,~T.; Morita,~A.; Ouchi,~Y. \emph{Phys Chem Chem Phys} \textbf{2008},
  \emph{10}, 2010--2013\relax
\mciteBstWouldAddEndPuncttrue
\mciteSetBstMidEndSepPunct{\mcitedefaultmidpunct}
{\mcitedefaultendpunct}{\mcitedefaultseppunct}\relax
\EndOfBibitem
\bibitem[Baer \latin{et~al.}(2009)Baer, Kuo, Bluhm, and Ghosal]{Ghosal09}
Baer,~M.~D.; Kuo,~I.-F.~W.; Bluhm,~H.; Ghosal,~S. \emph{J Phys Chem B}
  \textbf{2009}, \emph{113}, 15843--15850\relax
\mciteBstWouldAddEndPuncttrue
\mciteSetBstMidEndSepPunct{\mcitedefaultmidpunct}
{\mcitedefaultendpunct}{\mcitedefaultseppunct}\relax
\EndOfBibitem
\bibitem[Hua \latin{et~al.}(2013)Hua, Verreault, and Allen]{Allen13}
Hua,~W.; Verreault,~D.; Allen,~H.~C. \emph{J Phys Chem Lett} \textbf{2013},
  \emph{4}, 4231--4236\relax
\mciteBstWouldAddEndPuncttrue
\mciteSetBstMidEndSepPunct{\mcitedefaultmidpunct}
{\mcitedefaultendpunct}{\mcitedefaultseppunct}\relax
\EndOfBibitem
\bibitem[Tobias \latin{et~al.}(2013)Tobias, Stern, Baer, Levin, and
  Mundy]{Christopher13}
Tobias,~D.~J.; Stern,~A.~C.; Baer,~M.~D.; Levin,~Y.; Mundy,~C.~J. \emph{Annu
  Rev Phys Chem} \textbf{2013}, \emph{64}, 339--359\relax
\mciteBstWouldAddEndPuncttrue
\mciteSetBstMidEndSepPunct{\mcitedefaultmidpunct}
{\mcitedefaultendpunct}{\mcitedefaultseppunct}\relax
\EndOfBibitem
\bibitem[Brandes \latin{et~al.}(2014)Brandes, Karageorgiev, Viswanath, and
  Motschmann]{Motschmann14}
Brandes,~E.; Karageorgiev,~P.; Viswanath,~P.; Motschmann,~H. \emph{J Phys Chem
  C} \textbf{2014}, \emph{118}, 26629--26633\relax
\mciteBstWouldAddEndPuncttrue
\mciteSetBstMidEndSepPunct{\mcitedefaultmidpunct}
{\mcitedefaultendpunct}{\mcitedefaultseppunct}\relax
\EndOfBibitem
\bibitem[Hua \latin{et~al.}(2015)Hua, Verreault, and Allen]{Allen15}
Hua,~W.; Verreault,~D.; Allen,~H.~C. \emph{J Am Chem Soc} \textbf{2015},
  \emph{137}, 13920--13926\relax
\mciteBstWouldAddEndPuncttrue
\mciteSetBstMidEndSepPunct{\mcitedefaultmidpunct}
{\mcitedefaultendpunct}{\mcitedefaultseppunct}\relax
\EndOfBibitem
\bibitem[Heydweiller(1910)]{hey10}
Heydweiller,~A. \emph{Ann Phys} \textbf{1910}, \emph{338}, 145--185\relax
\mciteBstWouldAddEndPuncttrue
\mciteSetBstMidEndSepPunct{\mcitedefaultmidpunct}
{\mcitedefaultendpunct}{\mcitedefaultseppunct}\relax
\EndOfBibitem
\bibitem[Randles(1957)]{Randles57}
Randles,~J. \emph{Discuss Faraday Soc} \textbf{1957}, \emph{24}, 194--199\relax
\mciteBstWouldAddEndPuncttrue
\mciteSetBstMidEndSepPunct{\mcitedefaultmidpunct}
{\mcitedefaultendpunct}{\mcitedefaultseppunct}\relax
\EndOfBibitem
\bibitem[Randles and Schiffrin(1966)Randles, and Schiffrin]{ran66}
Randles,~J. E.~B.; Schiffrin,~D.~J. \emph{T Faraday Soc} \textbf{1966},
  \emph{62}, 2403--2408\relax
\mciteBstWouldAddEndPuncttrue
\mciteSetBstMidEndSepPunct{\mcitedefaultmidpunct}
{\mcitedefaultendpunct}{\mcitedefaultseppunct}\relax
\EndOfBibitem
\bibitem[Matubayasi \latin{et~al.}(1999)Matubayasi, Matsuo, Yamamoto,
  Yamaguchi, and Matuzawa]{mat99}
Matubayasi,~N.; Matsuo,~H.; Yamamoto,~K.; Yamaguchi,~S.; Matuzawa,~A. \emph{J
  Colloid Interface Sci} \textbf{1999}, \emph{209}, 398--402\relax
\mciteBstWouldAddEndPuncttrue
\mciteSetBstMidEndSepPunct{\mcitedefaultmidpunct}
{\mcitedefaultendpunct}{\mcitedefaultseppunct}\relax
\EndOfBibitem
\bibitem[Matubayasi \latin{et~al.}(2001)Matubayasi, Tsunetomo, Sato, Akizuki,
  Morishita, Matuzawa, and Natsukari]{mat01}
Matubayasi,~N.; Tsunetomo,~K.; Sato,~I.; Akizuki,~R.; Morishita,~T.;
  Matuzawa,~A.; Natsukari,~Y. \emph{J Colloid Interface Sci} \textbf{2001},
  \emph{243}, 444--456\relax
\mciteBstWouldAddEndPuncttrue
\mciteSetBstMidEndSepPunct{\mcitedefaultmidpunct}
{\mcitedefaultendpunct}{\mcitedefaultseppunct}\relax
\EndOfBibitem
\bibitem[Wagner(1924)]{Wagner24}
Wagner,~C. \emph{{Phys Z}} \textbf{1924}, \emph{25}, 474--477\relax
\mciteBstWouldAddEndPuncttrue
\mciteSetBstMidEndSepPunct{\mcitedefaultmidpunct}
{\mcitedefaultendpunct}{\mcitedefaultseppunct}\relax
\EndOfBibitem
\bibitem[Onsager and Samaras(1934)Onsager, and Samaras]{Samaras34}
Onsager,~L.; Samaras,~N. N.~T. \emph{J Chem Phys} \textbf{1934}, \emph{2},
  528--536\relax
\mciteBstWouldAddEndPuncttrue
\mciteSetBstMidEndSepPunct{\mcitedefaultmidpunct}
{\mcitedefaultendpunct}{\mcitedefaultseppunct}\relax
\EndOfBibitem
\bibitem[Knipping \latin{et~al.}(2000)Knipping, Lakin, Foster, Jungwirth,
  Tobias, Gerber, Dabdub, and Finlayson-Pitts]{Knipping301}
Knipping,~E.~M.; Lakin,~M.~J.; Foster,~K.~L.; Jungwirth,~P.; Tobias,~D.~J.;
  Gerber,~R.~B.; Dabdub,~D.; Finlayson-Pitts,~B.~J. \emph{Science}
  \textbf{2000}, \emph{288}, 301--306\relax
\mciteBstWouldAddEndPuncttrue
\mciteSetBstMidEndSepPunct{\mcitedefaultmidpunct}
{\mcitedefaultendpunct}{\mcitedefaultseppunct}\relax
\EndOfBibitem
\bibitem[Padmanabhan \latin{et~al.}(2007)Padmanabhan, Daillant, Belloni, Mora,
  Alba, and Konovalov]{Padm07}
Padmanabhan,~V.; Daillant,~J.; Belloni,~L.; Mora,~S.; Alba,~M.; Konovalov,~O.
  \emph{Phys Rev Lett} \textbf{2007}, \emph{99}, 086105\relax
\mciteBstWouldAddEndPuncttrue
\mciteSetBstMidEndSepPunct{\mcitedefaultmidpunct}
{\mcitedefaultendpunct}{\mcitedefaultseppunct}\relax
\EndOfBibitem
\bibitem[Otten \latin{et~al.}(2012)Otten, Shaffer, Geissler, and
  Saykally]{Otten12}
Otten,~D.~E.; Shaffer,~P.~R.; Geissler,~P.~L.; Saykally,~R.~J. \emph{P Natl
  Acad Sci USA} \textbf{2012}, \emph{109}, 701--705\relax
\mciteBstWouldAddEndPuncttrue
\mciteSetBstMidEndSepPunct{\mcitedefaultmidpunct}
{\mcitedefaultendpunct}{\mcitedefaultseppunct}\relax
\EndOfBibitem
\bibitem[Bostr\"{o}m \latin{et~al.}(2001)Bostr\"{o}m, Williams, and
  Ninham]{Barry01}
Bostr\"{o}m,~M.; Williams,~D. R.~M.; Ninham,~B.~W. \emph{Langmuir}
  \textbf{2001}, \emph{17}, 4475--4478\relax
\mciteBstWouldAddEndPuncttrue
\mciteSetBstMidEndSepPunct{\mcitedefaultmidpunct}
{\mcitedefaultendpunct}{\mcitedefaultseppunct}\relax
\EndOfBibitem
\bibitem[Levin(2009)]{Levin09}
Levin,~Y. \emph{Phys Rev Lett} \textbf{2009}, \emph{102}, 147803\relax
\mciteBstWouldAddEndPuncttrue
\mciteSetBstMidEndSepPunct{\mcitedefaultmidpunct}
{\mcitedefaultendpunct}{\mcitedefaultseppunct}\relax
\EndOfBibitem
\bibitem[Levin \latin{et~al.}(2009)Levin, dos Santos, and Diehl]{Levin009}
Levin,~Y.; dos Santos,~A.~P.; Diehl,~A. \emph{Phys Rev Lett} \textbf{2009},
  \emph{103}, 257802\relax
\mciteBstWouldAddEndPuncttrue
\mciteSetBstMidEndSepPunct{\mcitedefaultmidpunct}
{\mcitedefaultendpunct}{\mcitedefaultseppunct}\relax
\EndOfBibitem
\bibitem[Wang and Wang(2014)Wang, and Wang]{Wang14}
Wang,~R.; Wang,~Z.-G. \emph{Phys Rev Lett} \textbf{2014}, \emph{112},
  136101\relax
\mciteBstWouldAddEndPuncttrue
\mciteSetBstMidEndSepPunct{\mcitedefaultmidpunct}
{\mcitedefaultendpunct}{\mcitedefaultseppunct}\relax
\EndOfBibitem
\bibitem[Netz and Horinek(2012)Netz, and Horinek]{net12}
Netz,~R.~R.; Horinek,~D. \emph{Annu Rev Phys Chem} \textbf{2012}, \emph{63},
  401--418\relax
\mciteBstWouldAddEndPuncttrue
\mciteSetBstMidEndSepPunct{\mcitedefaultmidpunct}
{\mcitedefaultendpunct}{\mcitedefaultseppunct}\relax
\EndOfBibitem
\bibitem[Caleman \latin{et~al.}(2011)Caleman, Hub, van Maaren, and van~der
  Spoel]{Caleman11}
Caleman,~C.; Hub,~J.~S.; van Maaren,~P.~J.; van~der Spoel,~D. \emph{P Natl Acad
  Sci USA} \textbf{2011}, \emph{108}, 6838--6842\relax
\mciteBstWouldAddEndPuncttrue
\mciteSetBstMidEndSepPunct{\mcitedefaultmidpunct}
{\mcitedefaultendpunct}{\mcitedefaultseppunct}\relax
\EndOfBibitem
\bibitem[Baer and Mundy(2011)Baer, and Mundy]{Baer11}
Baer,~M.~D.; Mundy,~C.~J. \emph{J Phys Chem Lett} \textbf{2011}, \emph{2},
  1088--1093\relax
\mciteBstWouldAddEndPuncttrue
\mciteSetBstMidEndSepPunct{\mcitedefaultmidpunct}
{\mcitedefaultendpunct}{\mcitedefaultseppunct}\relax
\EndOfBibitem
\bibitem[Baer \latin{et~al.}(2012)Baer, Stern, Levin, Tobias, and
  Mundy]{Baer12}
Baer,~M.~D.; Stern,~A.~C.; Levin,~Y.; Tobias,~D.~J.; Mundy,~C.~J. \emph{J Phys
  Chem Lett} \textbf{2012}, \emph{3}, 1565--1570\relax
\mciteBstWouldAddEndPuncttrue
\mciteSetBstMidEndSepPunct{\mcitedefaultmidpunct}
{\mcitedefaultendpunct}{\mcitedefaultseppunct}\relax
\EndOfBibitem
\bibitem[Sun \latin{et~al.}(2015)Sun, Li, Tu, and Agren]{Sun15}
Sun,~L.; Li,~X.; Tu,~Y.; Agren,~H. \emph{Phys Chem Chem Phys} \textbf{2015},
  \emph{17}, 4311--4318\relax
\mciteBstWouldAddEndPuncttrue
\mciteSetBstMidEndSepPunct{\mcitedefaultmidpunct}
{\mcitedefaultendpunct}{\mcitedefaultseppunct}\relax
\EndOfBibitem
\bibitem[Yang \latin{et~al.}(2001)Yang, Dabros, Li, Czarnecki, and
  Masliyah]{yan01}
Yang,~C.; Dabros,~T.; Li,~D.; Czarnecki,~J.; Masliyah,~J.~H. \emph{J Colloid
  Interface Sci} \textbf{2001}, \emph{243}, 128--135\relax
\mciteBstWouldAddEndPuncttrue
\mciteSetBstMidEndSepPunct{\mcitedefaultmidpunct}
{\mcitedefaultendpunct}{\mcitedefaultseppunct}\relax
\EndOfBibitem
\bibitem[Hirose \latin{et~al.}(1992)Hirose, Akamatsu, and Domen]{hir92}
Hirose,~C.; Akamatsu,~N.; Domen,~K. \emph{J Chem Phys} \textbf{1992},
  \emph{95}, 997--1004\relax
\mciteBstWouldAddEndPuncttrue
\mciteSetBstMidEndSepPunct{\mcitedefaultmidpunct}
{\mcitedefaultendpunct}{\mcitedefaultseppunct}\relax
\EndOfBibitem
\bibitem[Hamm(2014)]{ham14c}
Hamm,~P. \emph{J Chem Phys} \textbf{2014}, \emph{141}, 184201\relax
\mciteBstWouldAddEndPuncttrue
\mciteSetBstMidEndSepPunct{\mcitedefaultmidpunct}
{\mcitedefaultendpunct}{\mcitedefaultseppunct}\relax
\EndOfBibitem
\bibitem[Ito \latin{et~al.}(2016)Ito, Hasegawa, and Tanimura]{ito16}
Ito,~H.; Hasegawa,~T.; Tanimura,~Y. \emph{J Phys Chem Lett} \textbf{2016},
  \emph{7}, 4147--4151\relax
\mciteBstWouldAddEndPuncttrue
\mciteSetBstMidEndSepPunct{\mcitedefaultmidpunct}
{\mcitedefaultendpunct}{\mcitedefaultseppunct}\relax
\EndOfBibitem
\bibitem[Lambert \latin{et~al.}(2005)Lambert, Davies, and Neivandt]{lam05}
Lambert,~A.~G.; Davies,~P.~B.; Neivandt,~D.~J. \emph{Appl Spect Rev}
  \textbf{2005}, \emph{40}, 103--145\relax
\mciteBstWouldAddEndPuncttrue
\mciteSetBstMidEndSepPunct{\mcitedefaultmidpunct}
{\mcitedefaultendpunct}{\mcitedefaultseppunct}\relax
\EndOfBibitem
\bibitem[Ratcliffe and Irish(1984)Ratcliffe, and Irish]{Irish1984}
Ratcliffe,~C.; Irish,~D. \emph{Can J Chem} \textbf{1984}, \emph{62},
  1134--1144\relax
\mciteBstWouldAddEndPuncttrue
\mciteSetBstMidEndSepPunct{\mcitedefaultmidpunct}
{\mcitedefaultendpunct}{\mcitedefaultseppunct}\relax
\EndOfBibitem
\bibitem[Marcus(2016)]{mar16c}
Marcus,~Y. \emph{Current Opinion In Colloid {\&} Interface Science}
  \textbf{2016}, \emph{23}, 94--99\relax
\mciteBstWouldAddEndPuncttrue
\mciteSetBstMidEndSepPunct{\mcitedefaultmidpunct}
{\mcitedefaultendpunct}{\mcitedefaultseppunct}\relax
\EndOfBibitem
\bibitem[Tong \latin{et~al.}(2016)Tong, Kampfrath, and Campen]{tong16PCCP}
Tong,~Y.; Kampfrath,~T.; Campen,~R.~K. \emph{Phys Chem Chem Phys}
  \textbf{2016}, \emph{18}, 18424--18430\relax
\mciteBstWouldAddEndPuncttrue
\mciteSetBstMidEndSepPunct{\mcitedefaultmidpunct}
{\mcitedefaultendpunct}{\mcitedefaultseppunct}\relax
\EndOfBibitem
\bibitem[Hyodo(1989)]{Hyodo1989}
Hyodo,~S.-a. \emph{Chem Phys Lett} \textbf{1989}, \emph{161}, 245--248\relax
\mciteBstWouldAddEndPuncttrue
\mciteSetBstMidEndSepPunct{\mcitedefaultmidpunct}
{\mcitedefaultendpunct}{\mcitedefaultseppunct}\relax
\EndOfBibitem
\bibitem[Karelin and Grigorovich(1976)Karelin, and
  Grigorovich]{Grigorovich1976}
Karelin,~A.; Grigorovich,~Z. \emph{Spectrochim Acta A} \textbf{1976},
  \emph{32}, 851--857\relax
\mciteBstWouldAddEndPuncttrue
\mciteSetBstMidEndSepPunct{\mcitedefaultmidpunct}
{\mcitedefaultendpunct}{\mcitedefaultseppunct}\relax
\EndOfBibitem
\bibitem[Baer \latin{et~al.}(2014)Baer, Tobias, and Mundy]{Baer12a}
Baer,~M.~D.; Tobias,~D.~J.; Mundy,~C.~J. \emph{J Phys Chem C} \textbf{2014},
  \emph{118}, 29412--29420\relax
\mciteBstWouldAddEndPuncttrue
\mciteSetBstMidEndSepPunct{\mcitedefaultmidpunct}
{\mcitedefaultendpunct}{\mcitedefaultseppunct}\relax
\EndOfBibitem
\bibitem[Chen \latin{et~al.}(2004)Chen, Zhang, and Zhao]{che04b}
Chen,~Y.; Zhang,~Y.-H.; Zhao,~L.-J. \emph{Phys Chem Chem Phys} \textbf{2004},
  \emph{6}, 537--542\relax
\mciteBstWouldAddEndPuncttrue
\mciteSetBstMidEndSepPunct{\mcitedefaultmidpunct}
{\mcitedefaultendpunct}{\mcitedefaultseppunct}\relax
\EndOfBibitem
\bibitem[Ritzhaupt and Devlin(1975)Ritzhaupt, and Devlin]{rit75}
Ritzhaupt,~G.; Devlin,~J.~P. \emph{J Chem Phys} \textbf{1975}, \emph{62},
  1982--1986\relax
\mciteBstWouldAddEndPuncttrue
\mciteSetBstMidEndSepPunct{\mcitedefaultmidpunct}
{\mcitedefaultendpunct}{\mcitedefaultseppunct}\relax
\EndOfBibitem
\bibitem[Ottosson \latin{et~al.}(2009)Ottosson, V\'{a}cha, Aziz, Pokapanich,
  Eberhardt, Svensson, \"{O}hrwall, Jungwirth, Bj\"{o}rneholm, and
  Winter]{ott09}
Ottosson,~N.; V\'{a}cha,~R.; Aziz,~E.~F.; Pokapanich,~W.; Eberhardt,~W.;
  Svensson,~S.; \"{O}hrwall,~G.; Jungwirth,~P.; Bj\"{o}rneholm,~O.; Winter,~B.
  \emph{The Journal of Chemical Physics} \textbf{2009}, \emph{131},
  124706\relax
\mciteBstWouldAddEndPuncttrue
\mciteSetBstMidEndSepPunct{\mcitedefaultmidpunct}
{\mcitedefaultendpunct}{\mcitedefaultseppunct}\relax
\EndOfBibitem
\bibitem[Hey \latin{et~al.}(2016)Hey, Smeeton, Oakley, and Johnston]{hey16}
Hey,~J.~C.; Smeeton,~L.~C.; Oakley,~M.~T.; Johnston,~R.~L. \emph{J Phys Chem A}
  \textbf{2016}, \emph{120}, 4008--4015\relax
\mciteBstWouldAddEndPuncttrue
\mciteSetBstMidEndSepPunct{\mcitedefaultmidpunct}
{\mcitedefaultendpunct}{\mcitedefaultseppunct}\relax
\EndOfBibitem
\bibitem[dos Santos \latin{et~al.}(2010)dos Santos, Diehl, and Levin]{Santos10}
dos Santos,~A.~P.; Diehl,~A.; Levin,~Y. \emph{Langmuir} \textbf{2010},
  \emph{26}, 10778--10783\relax
\mciteBstWouldAddEndPuncttrue
\mciteSetBstMidEndSepPunct{\mcitedefaultmidpunct}
{\mcitedefaultendpunct}{\mcitedefaultseppunct}\relax
\EndOfBibitem
\bibitem[Levin and {dos Santos}(2014)Levin, and {dos Santos}]{lev14}
Levin,~Y.; {dos Santos},~A.~P. \emph{J Phys: Cond Matt} \textbf{2014},
  \emph{26}, 203101\relax
\mciteBstWouldAddEndPuncttrue
\mciteSetBstMidEndSepPunct{\mcitedefaultmidpunct}
{\mcitedefaultendpunct}{\mcitedefaultseppunct}\relax
\EndOfBibitem
\bibitem[Kathmann \latin{et~al.}(2011)Kathmann, Kuo, Mundy, and
  Schenter]{kat11}
Kathmann,~S.~M.; Kuo,~I.-F.~W.; Mundy,~C.~J.; Schenter,~G.~K. \emph{J Phys Chem
  B} \textbf{2011}, \emph{115}, 4369--4377\relax
\mciteBstWouldAddEndPuncttrue
\mciteSetBstMidEndSepPunct{\mcitedefaultmidpunct}
{\mcitedefaultendpunct}{\mcitedefaultseppunct}\relax
\EndOfBibitem
\bibitem[Gopalakrishnan \latin{et~al.}(2006)Gopalakrishnan, Liu, Allen, Kuo,
  and Shultz]{gop06}
Gopalakrishnan,~S.; Liu,~D.; Allen,~H.~C.; Kuo,~M.; Shultz,~M.~J. \emph{Chem
  Rev} \textbf{2006}, \emph{106}, 1155--1175\relax
\mciteBstWouldAddEndPuncttrue
\mciteSetBstMidEndSepPunct{\mcitedefaultmidpunct}
{\mcitedefaultendpunct}{\mcitedefaultseppunct}\relax
\EndOfBibitem
\bibitem[Tong \latin{et~al.}(2015)Tong, Wirth, Kirsch, Wolf, Saalfrank, and
  Campen]{tong15}
Tong,~Y.; Wirth,~J.; Kirsch,~H.; Wolf,~M.; Saalfrank,~P.; Campen,~R.~K. \emph{J
  Chem Phys} \textbf{2015}, \emph{142}, 054704\relax
\mciteBstWouldAddEndPuncttrue
\mciteSetBstMidEndSepPunct{\mcitedefaultmidpunct}
{\mcitedefaultendpunct}{\mcitedefaultseppunct}\relax
\EndOfBibitem
\bibitem[Tong \latin{et~al.}(2013)Tong, Vila~Verde, and Campen]{tong13}
Tong,~Y.; Vila~Verde,~A.; Campen,~R.~K. \emph{J Phys Chem B} \textbf{2013},
  \emph{117}, 11753--11764\relax
\mciteBstWouldAddEndPuncttrue
\mciteSetBstMidEndSepPunct{\mcitedefaultmidpunct}
{\mcitedefaultendpunct}{\mcitedefaultseppunct}\relax
\EndOfBibitem
\bibitem[Bain \latin{et~al.}(1991)Bain, Davies, Ong, Ward, and Brown]{bain91}
Bain,~C.~D.; Davies,~P.~B.; Ong,~T.~H.; Ward,~R.~N.; Brown,~M.~A.
  \emph{Langmuir} \textbf{1991}, \emph{7}, 1563--1566\relax
\mciteBstWouldAddEndPuncttrue
\mciteSetBstMidEndSepPunct{\mcitedefaultmidpunct}
{\mcitedefaultendpunct}{\mcitedefaultseppunct}\relax
\EndOfBibitem
\bibitem[Zhuang \latin{et~al.}(1999)Zhuang, Miranda, Kim, and Shen]{zhuang99}
Zhuang,~X.; Miranda,~P.~B.; Kim,~D.; Shen,~Y.~R. \emph{Phys Rev B}
  \textbf{1999}, \emph{59}, 12632--12640\relax
\mciteBstWouldAddEndPuncttrue
\mciteSetBstMidEndSepPunct{\mcitedefaultmidpunct}
{\mcitedefaultendpunct}{\mcitedefaultseppunct}\relax
\EndOfBibitem
\end{mcitethebibliography}

\begin{mcitethebibliography}{11}
\providecommand*\natexlab[1]{#1}
\providecommand*\mciteSetBstSublistMode[1]{}
\providecommand*\mciteSetBstMaxWidthForm[2]{}
\providecommand*\mciteBstWouldAddEndPuncttrue
  {\def\EndOfBibitem{\unskip.}}
\providecommand*\mciteBstWouldAddEndPunctfalse
  {\let\EndOfBibitem\relax}
\providecommand*\mciteSetBstMidEndSepPunct[3]{}
\providecommand*\mciteSetBstSublistLabelBeginEnd[3]{}
\providecommand*\EndOfBibitem{}
\mciteSetBstSublistMode{f}
\mciteSetBstMaxWidthForm{subitem}{(\alph{mcitesubitemcount})}
\mciteSetBstSublistLabelBeginEnd
  {\mcitemaxwidthsubitemform\space}
  {\relax}
  {\relax}

\bibitem[Tong \latin{et~al.}(2016)Tong, Kampfrath, and Campen]{tong16PCCP}
Tong,~Y.; Kampfrath,~T.; Campen,~R.~K. \emph{Phys Chem Chem Phys}
  \textbf{2016}, \emph{18}, 18424--18430\relax
\mciteBstWouldAddEndPuncttrue
\mciteSetBstMidEndSepPunct{\mcitedefaultmidpunct}
{\mcitedefaultendpunct}{\mcitedefaultseppunct}\relax
\EndOfBibitem
\bibitem[Long(2001)]{lon01}
Long,~D.~A. \emph{The Raman Effect: A Unified Treatment of the Theory of Raman
  Scattering by Molecules}; Wiley, 2001\relax
\mciteBstWouldAddEndPuncttrue
\mciteSetBstMidEndSepPunct{\mcitedefaultmidpunct}
{\mcitedefaultendpunct}{\mcitedefaultseppunct}\relax
\EndOfBibitem
\bibitem[Hirose \latin{et~al.}(1992)Hirose, Akamatsu, and Domen]{hir92}
Hirose,~C.; Akamatsu,~N.; Domen,~K. \emph{J Chem Phys} \textbf{1992},
  \emph{95}, 997--1004\relax
\mciteBstWouldAddEndPuncttrue
\mciteSetBstMidEndSepPunct{\mcitedefaultmidpunct}
{\mcitedefaultendpunct}{\mcitedefaultseppunct}\relax
\EndOfBibitem
\bibitem[Wang \latin{et~al.}(2005)Wang, Gan, Lu, Rao, and Wu]{wan05}
Wang,~H.; Gan,~W.; Lu,~R.; Rao,~Y.; Wu,~B.-H. \emph{Int Rev Phys Chem}
  \textbf{2005}, \emph{24}, 191--256\relax
\mciteBstWouldAddEndPuncttrue
\mciteSetBstMidEndSepPunct{\mcitedefaultmidpunct}
{\mcitedefaultendpunct}{\mcitedefaultseppunct}\relax
\EndOfBibitem
\bibitem[Lambert \latin{et~al.}(2005)Lambert, Davies, and Neivandt]{lam05}
Lambert,~A.~G.; Davies,~P.~B.; Neivandt,~D.~J. \emph{Appl Spect Rev}
  \textbf{2005}, \emph{40}, 103--145\relax
\mciteBstWouldAddEndPuncttrue
\mciteSetBstMidEndSepPunct{\mcitedefaultmidpunct}
{\mcitedefaultendpunct}{\mcitedefaultseppunct}\relax
\EndOfBibitem
\bibitem[Hyodo(1989)]{Hyodo1989}
Hyodo,~S.-a. \emph{Chem Phys Lett} \textbf{1989}, \emph{161}, 245--248\relax
\mciteBstWouldAddEndPuncttrue
\mciteSetBstMidEndSepPunct{\mcitedefaultmidpunct}
{\mcitedefaultendpunct}{\mcitedefaultseppunct}\relax
\EndOfBibitem
\bibitem[Perdew \latin{et~al.}(1996)Perdew, Burke, and Ernzerhof]{per96}
Perdew,~J.; Burke,~K.; Ernzerhof,~M. \emph{Phys Rev Lett} \textbf{1996}, \relax
\mciteBstWouldAddEndPunctfalse
\mciteSetBstMidEndSepPunct{\mcitedefaultmidpunct}
{}{\mcitedefaultseppunct}\relax
\EndOfBibitem
\bibitem[Perdew \latin{et~al.}(1996)Perdew, Ernzerhof, and Burke]{per96b}
Perdew,~J.~P.; Ernzerhof,~M.; Burke,~K. \emph{J Chem Phys} \textbf{1996},
  \emph{105}, 9982--9985\relax
\mciteBstWouldAddEndPuncttrue
\mciteSetBstMidEndSepPunct{\mcitedefaultmidpunct}
{\mcitedefaultendpunct}{\mcitedefaultseppunct}\relax
\EndOfBibitem
\bibitem[Becke(1993)]{bec93}
Becke,~A.~D. \emph{J Chem Phys} \textbf{1993}, \emph{98}, 5648--5652\relax
\mciteBstWouldAddEndPuncttrue
\mciteSetBstMidEndSepPunct{\mcitedefaultmidpunct}
{\mcitedefaultendpunct}{\mcitedefaultseppunct}\relax
\EndOfBibitem
\bibitem[Schmidt \latin{et~al.}(1993)Schmidt, Baldridge, Boatz, Elbert, Gordon,
  Jensen, Koseki, Matsunaga, Nguyen, Su, Windus, Dupuis, and {Montgomery
  Jr}]{sch93b}
Schmidt,~M.~W.; Baldridge,~K.~K.; Boatz,~J.~A.; Elbert,~S.~T.; Gordon,~M.~S.;
  Jensen,~J.~H.; Koseki,~S.; Matsunaga,~N.; Nguyen,~K.~A.; Su,~S.;
  Windus,~T.~L.; Dupuis,~M.; {Montgomery Jr},~J.~A. \emph{J Comput Chem}
  \textbf{1993}, \emph{14}, 1347--1363\relax
\mciteBstWouldAddEndPuncttrue
\mciteSetBstMidEndSepPunct{\mcitedefaultmidpunct}
{\mcitedefaultendpunct}{\mcitedefaultseppunct}\relax
\EndOfBibitem
\end{mcitethebibliography}
%\defaultbibliography{REF_ClO4}

\providecommand{\latin}[1]{#1}
\providecommand*\mcitethebibliography{\thebibliography}
\csname @ifundefined\endcsname{endmcitethebibliography}
  {\let\endmcitethebibliography\endthebibliography}{}

\clearpage

\beginsupplement

\section{\textit{Electronic Supplementary Information for} Experimentally Quantifying Anion Polarizability at the Air/Water Interface}

\subsection{VSF Spectra of 0.6 M Perchlorate Salt Solutions}
\begin{figure}[htbp]
	\begin{center}
		\includegraphics[width=0.5\textwidth]{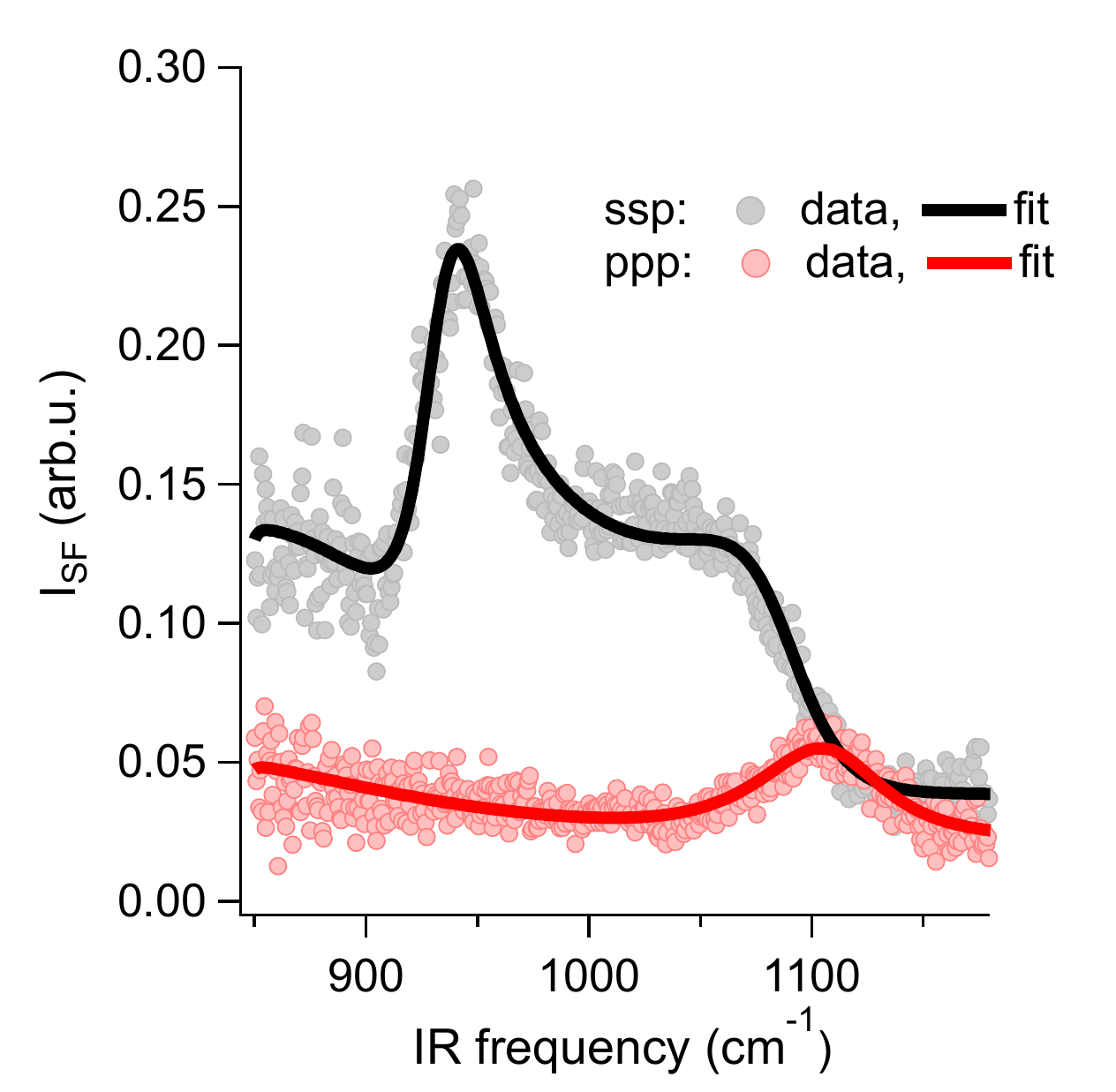}
		\caption{I$_{\text{sf}}$ spectrum of an 0.6 M \ce{NaClO4} solution plotted as a function of IR frequency. Clearly the spectrum, and the intensity of the $\nu_{1}$ mode, is quantitatively similar to that for an \ce{HClO4} solution (shown in the manuscript).}
		\label{f:salt}
	\end{center}
\end{figure}

I$_{\text{sf}}$ is plotted as a function of incident infrared frequency for an 0.6 M solution of \ce{NaClO4} in Figure \ref{f:salt}. Clearly this solution also has a distinct $\nu_{1}$ mode. Comparison with the spectra plotted in Figure 1 in the manuscript suggests that the break in symmetry that makes $\nu_{1}$ IR active near the interface is not the result of an interfacial change in pKa.

\subsection{Line Shape Analysis Details \& Full Results}

We do a line shape analysis of our experimental data by performing a global fit of 
spectra -- using the Levenberg-Marquardt algorithm as implemented in the 
commercial graphing and analysis program Igor Pro (Wavemetrics) -- collected under 
the \textit{ssp} and \textit{ppp} polarization conditions for all bulk \ce{HClO4} 
concentrations using the line shape expression described in the text. To quantitatively account for the effect of the, finite, visible pulse spectral width we additional convolve this response with a gaussian of width $\Delta\nu_{800}$. The spectrum of the visible pulse is independently measured before each VSF measurement. To do the fit we assume a libration (of interfacial water) whose 
center frequency and line width we have determined in our previous study 
\cite{tong16PCCP} and further assume, for concentrations lower than 1 M 
\ce{HClO4}, that the center frequencies and damping constants of the $\nu_{1}$ and 
$\nu_{3}$ modes; the libration amplitude; and the nonresonant amplitude and phase are all independent of concentration. Uncertainties reported with each 
fit parameter are calculated from a linearization of the model, with respect to 
its parameters, near the best fit. All results are shown in Tables \ref{t:ssp} and 
\ref{t:ppp}.
\begin{table}
	\begin{center}
	\scriptsize{
			\begin{tabular}{rlcccccccc}
%			\multicolumn{10}{c}{\normalsize{\textbf{\textit{ssp} polarization condition}}} \\
			\toprule
			{[HClO$_{4}$]} & (mol/L) &  0 & 0.1 & 0.2 & 0.3 & 0.4 & 0.5 & 0.6 & 0.8 \\ 
\midrule
			$\chi_{\text{nr}}$ &  &\multicolumn{8}{c}{$0.34 \pm 0.07$} \\    
			$\epsilon$  & (rad) & \multicolumn{8}{c}{$4.55\pm 0.13$} \\ 
			$\Delta\nu_{\text{800}}$ & (cm$^{\text{-1}}$) & \multicolumn{8}{c}{12.2}   \\
			\midrule 
			$\chi_{\nu_{\text{1}}}$ & & 0 & $1.32\pm0.1$ & $2.03\pm0.1$ & $2.41\pm0.2$ & $2.69\pm0.1$ & $2.73\pm0.1$ & $2.91\pm0.3$ & $3.23\pm0.2$ \\ 
			$\tilde{\nu}_{\nu_{1}}$ & (cm$^{\text{-1}}$) & \multicolumn{8}{c}{$935\pm1.0$}  \\ 
			$\Gamma_{\nu_{1}}$ & (cm$^{\text{-1}}$) & \multicolumn{8}{c}{$12.1\pm0.7$}  \\ 
			$\chi_{\nu_{\text{3}}}$ & & 0 & $-2.44\pm1.6$ & $-5.71\pm0.4$ & $-6.28\pm1.0$ & $-6.56\pm0.9$ & $-7.98\pm0.37$ & $-8.03\pm1.1$ & $-8.14\pm1.4$ \\ 
			$\tilde{\nu}_{\nu_{3}}$ & (cm$^{\text{-1}}$)& \multicolumn{8}{c}{$1110\pm1.4$} \\ 
			$\Gamma_{\nu_{3}}$ & (cm$^{\text{-1}}$) & \multicolumn{8}{c}{$35.7\pm1.2$}  \\
				\midrule  
			$\chi_{\text{lib}}$ & & \multicolumn{8}{c}{$17.83\pm4.2$} \\
			 $\tilde{\nu}_{\text{lib}}$ & (cm$^{\text{-1}}$) & \multicolumn{8}{c}{832}  \\ 
			 $\Gamma_{\text{lib}\nu_{3}}$ & (cm$^{\text{-1}}$)  & \multicolumn{8}{c}{135} \\
			 \bottomrule
		\end{tabular}}
		\caption{Results of fits to data collected employing the \textit{ssp} 
polarization condition. The parameters shown in this and Table \ref{t:ppp} are the 
result of a global fit to \textit{ppp} and \textit{ssp} spectra at all 
concentrations of \ce{HClO4}. Note that the spectral width of the 800 nm pulse was independently measured before each VSF measurement and the libration center frequency and line width were extracted from our previous work\cite{tong16PCCP}.}
		\label{t:ssp}
	\end{center}
\end{table}

\begin{table}
	\begin{center}
	\scriptsize{
			\begin{tabular}{rlcccccccc}
%			\multicolumn{10}{c}{\normalsize{\textbf{\textit{ssp} polarization condition}}} \\
			\toprule
			{[HClO$_{4}$]} & (mol/L) &  0 & 0.1 & 0.2 & 0.3 & 0.4 & 0.5 & 0.6 & 0.8 \\ 
\midrule
			$\chi_{\text{nr}}$ &  &\multicolumn{8}{c}{$0.34\pm0.05$} \\    
			$\epsilon$  & (rad) & \multicolumn{8}{c}{$4.55\pm0.15$} \\ 
			$\Delta\nu_{\text{800}}$ & (cm$^{\text{-1}}$) & \multicolumn{8}{c}{12.2}   \\
			\midrule 
			$\chi_{\nu_{\text{1}}}$ & & 0 & $0.04\pm0.1$ & $0.25\pm0.1$ & $0.26\pm0.1$ & $0.27\pm0.1$ & $0.25\pm0.1$ & $0.24\pm0.1$ & $0.26\pm0.1$ \\ 
			$\tilde{\nu}_{\nu_{1}}$ & (cm$^{\text{-1}}$) & \multicolumn{8}{c}{$935\pm2.3$}  \\ 
			$\Gamma_{\nu_{1}}$ & (cm$^{\text{-1}}$) & \multicolumn{8}{c}{$12.1\pm0.6$}  \\ 
			$\chi_{\nu_{\text{3}}}$ & & 0 & $-2.80\pm1.6$ & $-2.37\pm0.4$ & $-5.52\pm1.0$ & $-6.49\pm0.9$ & $-6.98\pm0.37$ & $-7.21\pm1.1$ & $-7.14\pm1.4$ \\ 
			$\tilde{\nu}_{\nu_{3}}$ & (cm$^{\text{-1}}$)& \multicolumn{8}{c}{$1110\pm0.7$} \\ 
			$\Gamma_{\nu_{3}}$ & (cm$^{\text{-1}}$) & \multicolumn{8}{c}{$35.7\pm0.9$}  \\
				\midrule  
			$\chi_{\text{lib}}$ & & \multicolumn{8}{c}{$17.83\pm0.7$} \\
			 $\tilde{\nu}_{\text{lib}}$ & (cm$^{\text{-1}}$) & \multicolumn{8}{c}{832}  \\ 
			 $\Gamma_{\text{lib}\nu_{3}}$ & (cm$^{\text{-1}}$)  & \multicolumn{8}{c}{135} \\
			 \bottomrule
		\end{tabular}}
		\caption{Results of fits to data collected employing the \textit{ppp} 
polarization condition. The parameters shown in this and Table \ref{t:ssp} are the 
results of a global fit to \textit{ppp} and \textit{ssp} spectra at all 
concentrations of \ce{HClO4}. Note that the spectral width of the 800 nm pulse was independently measured before each VSF measurement and the libration center frequency and line width were extracted from our previous work\cite{tong16PCCP}.}
		\label{t:ppp}
	\end{center}
\end{table}

\begin{figure}[htbp]
	\begin{center}
				\includegraphics[width=0.5\textwidth]{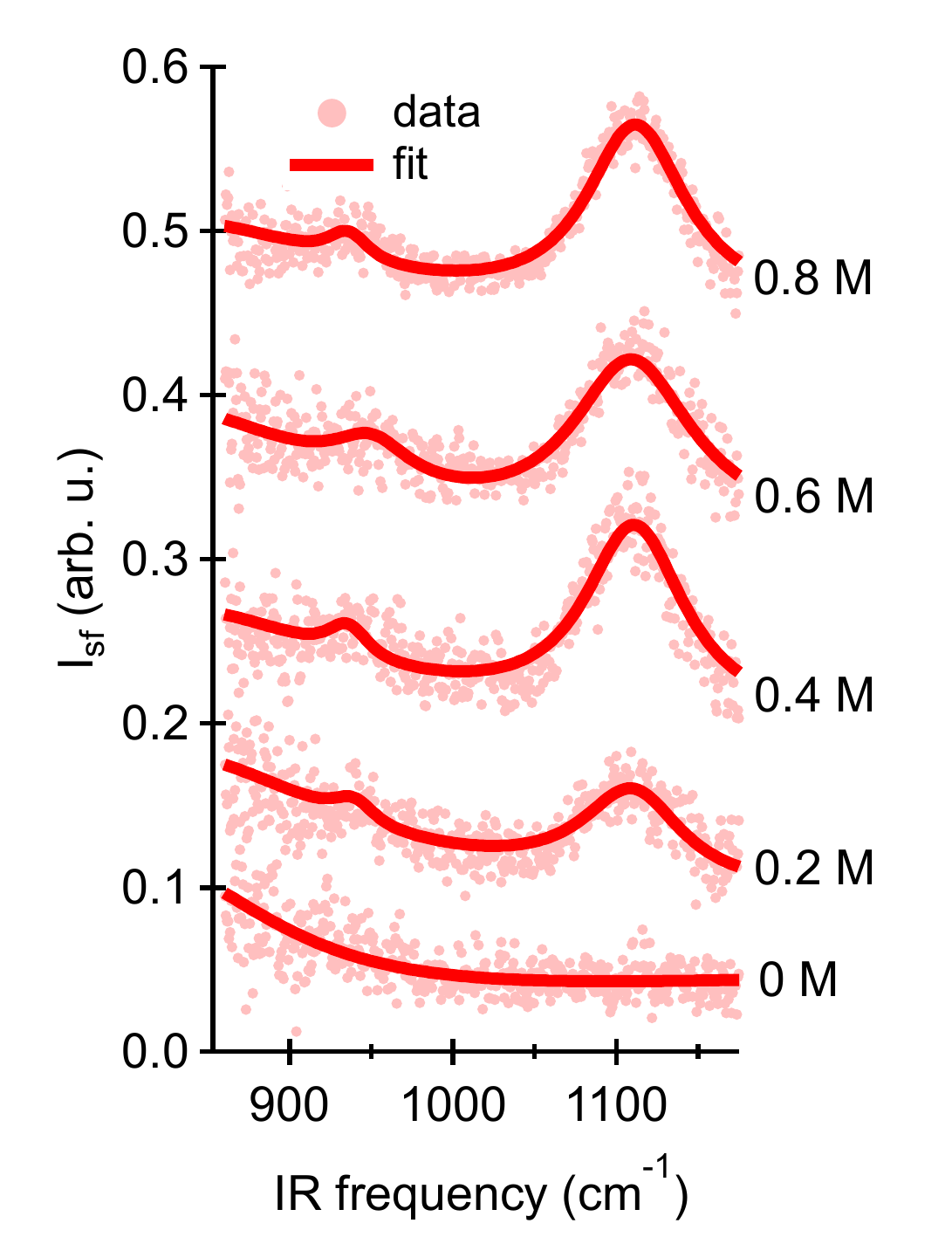}
			\caption{VSF spectra plotted as a function of bulk concentration of \ce{HClO4} collected under the \textit{ppp} polarization condition. Lines shown on the data are the results of global fits including the \textit{ssp} data shown in Figure 2 in the manuscript.}
		\label{f:ppp}
	\end{center}
\end{figure}

The \textit{ppp} spectra corresponding to the \textit{ssp} spectra shown in Figure 2 in the manuscript are shown in Figure \ref{f:ppp}. Clearly this signal is weak. However, we are interested in extracting the amplitude of the $\nu_{1}$ mode subject to the constraints described above. Given these boundary conditions we find, as is hopefully clear from inspection of the data, that a nonzero \textit{ppp} amplitude exists at all \ce{HClO4} concentrations 0.2 M and above. While we globally fit all data sets to extract the center frequencies of $\nu_{1}$ and $\nu_{3}$ the relatively weak $\nu_{1}$ amplitude, differing noise at the low frequency side of the measurement, and the coherent nature of the VSF response (leading to interference with the $\nu_{3}$), leads to small frequency shifts in the apparent peak in the signal even as the data is well described by a constant resonance frequency.  

\subsection{The Raman Depolarization Ratio in Bulk and at the Interface}
In the following section we develop the full description of the connection between 
the Raman depolarization ratio and measured VSF intensities as described in the 
literature by Long and Hirose et al \cite{lon01,hir92} and recently reviewed by 
Wang et al \cite{wan05}. Given a \ce{ClO4-} ion has C$_{3\nu}$ symmetry, 
assuming off diagonal terms in the polarizability tensor are small and the c-axis 
is taken along the rotational symmetry axis of the ion, there are three nonzero and 
two independent terms in the polarizability tensor: $\alpha_{aa} = \alpha_{bb}$, $
\alpha_{cc}$. For this molecule the Raman depolarization ratio is defined (where 
$R=\nicefrac{\alpha_{aa}}{\alpha_{cc}}$):
%%%%%
\begin{equation}\label{e:rho}
\rho = \frac{I_{\perp}}{I_{\parallel}} = \frac{3(\alpha_{a})^{2}}{45(\alpha_{i})^{2} + 4(\alpha_{a})^{2}} = \frac{3}{4 + 5\left[(1+2R)/(R-1)   \right]^{2} }
\end{equation}
$\alpha_{i}$ is defined, 
%%%%%
\begin{equation}\label{e:ai}
\alpha_{i} = \nicefrac{1}{3}\left(\alpha_{aa} + \alpha_{bb} + \alpha_{cc} \right)
\end{equation}
and $\alpha_{a}$ is defined,
%%%%%
\begin{equation}\label{e:aa}
\left(\alpha_{a} \right)^{2} = \nicefrac{1}{2}\left[(\alpha_{aa} - \alpha_{bb})^{2} + (\alpha_{bb} - \alpha_{cc})^{2}  + (\alpha_{cc} - \alpha_{aa})^{2} + 6(\alpha_{ab}^{2} + \alpha_{bc}^{2} + \alpha_{ca}^{2})\right]
\end{equation}
Clearly, then, if we could extract $R$ with interfacial specificity we could 
define an \textit{interfacial} Raman depolarization ratio.

As has been described extensively in the literature \cite{lam05} the measured VSF 
intensity, \textit{i.e.}\ $\text{I}_{\text{sf}}$, collected in reflection at the 
air/water interface can be written:
%%%
\begin{equation}\label{e:inten}
\text{I}_{\text{sf}} = \frac{8\pi^{3}\omega_{\text{sf}}^{2
}\sec^{2}\gamma_{\text{sf}}}{c^{3}}\left|\chi^{(2)}_{\text{eff}} \right|^{2}\text{I}_{\text{vis}}\text{I}_{\text{ir}}
\end{equation}
%%%
in which $\gamma_{i}$ is the angle of beam $i$ with respect to the surface normal, 
$\omega_{i}$ is the frequency of field $i$, $c$ is the speed of light, I$_{i}$ is 
the intensity of field $i$ and $\chi^{(2)}_{\text{eff}}$ is the effective, 
macroscopic, nonlinear susceptibility of the air/water interface. $\chi^{(2)}
_{\text{eff}}$ is a function of the nonlinear Fresnel factors ($L_{ij}$) and the 
polarizations of the incident and outgoing fields. These relationships can be 
written (assuming $z$ is along the surface normal and $x,y$ the plane of the 
surface), 
%%%
\begin{eqnarray}
\chi^{(2)}_{\text{eff,}ssp} & = & L_{yy}(\omega_{\text{sf}})L_{yy}(\omega_{\text{vis}})L_{zz}(\omega_{\text{ir}})\sin\gamma_{\text{ir}}\chi^{(2)}_{yyz} \label{e:ssp}\\
\chi^{(2)}_{\text{eff,}ppp} & = & -L_{xx}(\omega_{\text{sf}})L_{xx}(\omega_{\text{vis}})L_{zz}(\omega_{\text{ir}})\cos\gamma_{\text{sf}}\cos\gamma_{\text{vis}}\sin\gamma_{\text{ir}}\chi^{(2)}_{xxz}  \nonumber\\
  &  & -L_{xx}(\omega_{\text{sf}})L_{zz}(\omega_{\text{vis}})L_{xx}(\omega_{\text{ir}})\cos\gamma_{\text{sf}}\sin\gamma_{\text{vis}}\cos\gamma_{\text{ir}}\chi^{(2)}_{xzx} \label{e:ppp} \\
  &  & +L_{zz}(\omega_{\text{sf}})L_{xx}(\omega_{\text{vis}})L_{xx}(\omega_{\text{ir}})\sin\gamma_{\text{sf}}\cos\gamma_{\text{vis}}\cos\gamma_{\text{ir}}\chi^{(2)}_{zxx} \nonumber \\
  &  & +L_{zz}(\omega_{\text{sf}})L_{zz}(\omega_{\text{vis}})L_{zz}(\omega_{\text{ir}})\sin\gamma_{\text{sf}}\sin\gamma_{\text{vis}}\sin\gamma_{\text{ir}}\chi^{(2)}_{zzz} \nonumber
\end{eqnarray}
%%%
The nonlinear Fresnel factors are a function of the bulk and interfacial refractive indices and beam angles,
\begin{eqnarray}
	L_{xx}(\omega_{i}) & = &  \frac{2n_{\text{air}}(\omega_{i})\cos\zeta_{i}}{n_{\text{air}}(\omega_{i})\cos\zeta_{i} + n_{\text{water}}(\omega_i)\cos\gamma_{i}} \label{e:xx}\\
	L_{yy}(\omega_{i}) & = &  \frac{2n_{\text{air}}(\omega_{i})\cos\gamma_{i}}{n_{\text{air}}(\omega_{i})\cos\gamma_{i} + n_{\text{water}}(\omega_i)\cos\zeta_{i}} \label{e:yy}\\
	L_{zz}(\omega_{i}) & = & \frac{2n_{\text{water}}(\omega_{i})\cos\gamma_{i}}{n_{\text{air}}(\omega_{i})\cos\beta_{i} + n_{\text{water}}(\omega_i)\cos\gamma_{i}}\left(\frac{n_{\text{air}}(\omega_{i})}{n^{\prime}(\omega_{i})} \right)^{2} \label{e:zz}
\end{eqnarray}
in which $\zeta_{i}$ is the refracted angle of beam $i$ (\textit{i.e.}\ 
$n_{\text{air}}(\omega_{i})\sin\gamma_{i} = n_{\text{water}}(\omega_{i})
\sin\zeta_{i}$), $n_{i}$ is the, frequency dependent, refractive index of bulk 
phase $i$, and $n^{\prime}$ is the, also frequency dependent, refractive index of 
the interface. The material nonlinear susceptibility in the lab frame, 
\textit{i.e.}\ $\chi^{(2)}_{ijk}$, can be expressed in terms of the nonlinear 
molecular response, and the ensemble averaged orientation of ions with a C$_{3\nu}$ symmetry symmetric stretch as:
\begin{eqnarray}
\chi^{(2)}_{zzz} & = & N_{s}\beta^{(2)}_{ccc}\left[R\langle \cos\theta \rangle + \langle\cos^{3}\theta\rangle(1-R) \right] \label{e:zzz}\\
\chi^{(2)}_{xxz} & = & \chi^{(2)}_{yyz} = \nicefrac{1}{2}N_{s}\beta^{(2)}_{ccc}\left[\langle\cos\theta\rangle(1+R) - \langle\cos^{3}\theta\rangle(1-R) \right] \label{e:xxz}\\
\chi^{(2)}_{xzx} & = & \chi^{(2)}_{yzy} = \chi^{(2)}_{zxx} = \chi^{(2)}_{zyy} \nonumber \label{e:xzx} \\
				 & = & \nicefrac{1}{2}N_{s}\beta^{(2)}_{ccc}(1-R)\left[\langle\cos\theta\rangle - \langle\cos^{3}\theta\rangle \right]
\end{eqnarray}
in which $\beta^{(2)}_{abc}$ is the hyperpolarizability (\textit{i.e.}\ the 
molecular nonlinear response), the c-axis is the rotational symmetry of the 
C$_{3\nu}$ molecule, $\theta$ is the orientation of the \ce{ClO4-} with respect to 
the surface normal (the z-axis) and $R=\nicefrac{\beta^{(2)}_{aac}}{\beta^{(2)}
_{ccc}}$.

In this study we employed incident beams in the visible and infrared. These 
frequencies were chosen such that the infrared is resonant with Cl-O vibrations 
and the visible is nonresonant. Under such conditions $\beta^{(2)}$ is an anti-stokes 
scattering from an IR induced polarization:
%%%
\begin{equation}\label{e:beta}
	\beta^{(2)}_{ijk} = \frac{1}{2\hbar}\frac{\alpha_{ij}\mu_{k}}{\left(\omega_{n} - \omega_{\text{ir}} - i\Gamma_{n} \right)}
\end{equation}
%%%
in which $\hbar$ is the reduced Planck's constant, $\omega_{n}$ is the center 
frequency of the $n^{th}$ vibration, $\omega_{\text{ir}}$ is the frequency of the 
incident ir,  $\Gamma_{n}$ is the damping constant of the $n^{th}$ mode, $
\alpha_{ij}$ is the polarizability tensor (as described in equations \ref{e:ai}-\ref{e:aa}) and $\mu_{k}$ is the transition dipole. Given equation \ref{e:beta}, 
substituting equations \ref{e:zzz}, \ref{e:xxz} and \ref{e:xzx} and equations 
\ref{e:xx}, \ref{e:yy} and \ref{e:zz} into equations \ref{e:ssp} and \ref{e:ppp} 
suggests that, if we know the orientation of the \ce{ClO4-} and measure 
I$_{\text{sf}}$ under the \textit{ppp} and \textit{ssp} polarization conditions, 
the $\nicefrac{ppp}{ssp}$ ratio depends only on $R$. Because $\beta^{(2)}$ is a 
product of the polarizability and transition dipole:
%%%
\begin{equation}\label{e:R}
	R = \frac{\beta^{(2)}_{aac}}{\beta^{(2)}_{ccc}} = \frac{\alpha_{aa}\times\mu_{c}}{\alpha_{cc}\times\mu_{c}} = \frac{\alpha_{aa}}{\alpha_{cc}}
\end{equation}
%%% 
Equation \ref{e:R} thus implies that by taking the ratio of $\text{I}_{\text{sf}}$ measured under the \textit{ppp} and \textit{ssp} polarization conditions we can extract an $R$ for interfacial \ce{ClO4-} and calculate an interfacial Raman depolarization ratio.

\subsection{Evaluating Assumptions in the Calculation of Interfacial $\rho$}
%%%%
\subsubsection{What is the dependence of calculated $\rho$ on ClO$_{4}^{-}$ orientation?}
\begin{figure}[htbp]
	\begin{center}
			\includegraphics[width=0.7\textwidth]{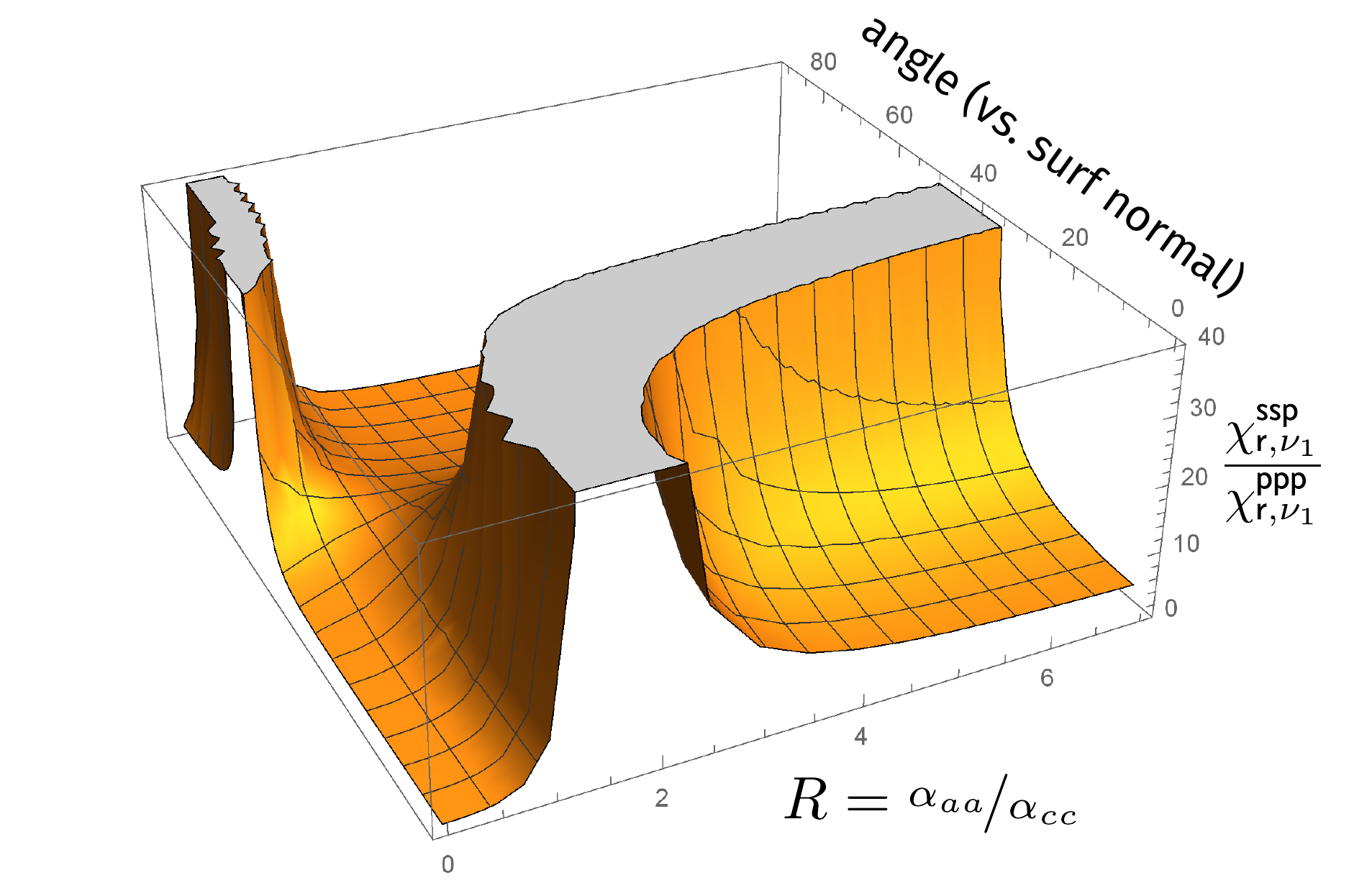}
		\caption{Calculated relationship of the $R$ at the IR frequencies of the $
\nu_{1}$ mode, measured  $\nicefrac{\chi^{\mathsf{ssp}}_{\mathsf{r},\nu_{1}}}
{\chi^{\mathsf{ppp}}_{\mathsf{r},\nu_{1}}}$ and \ce{ClO4-} orientation. The 
physically relevant solution to this set of equations is that with an $R$ value 
between 0 and 1.}
		\label{f:sens}
	\end{center}
\end{figure}

Given the set of equations shown, an experimentally measured $
\nicefrac{\chi^{\mathsf{ssp}}_{\mathsf{r}}}{\chi^{\mathsf{ppp}}_{\mathsf{r}}}$ 
ratio, and assuming a \ce{ClO4-} orientation one can extract a value of $R$. 
The resulting solution is plotted in Figure \ref{f:sens}. Calculated interfacial 
values of $\rho$ shown in the manuscript assume the \ce{ClO4-} anion is oriented 
between 0 and 45$^{\circ}$ and $R$ values range between 0 and 1. Orientations 
between 45 and 90 degrees would imply ion pairing and (further) reduced \ce{ClO4-} 
symmetry for which we see no spectral evidence. $R$ values above 2 imply that the 
$\nu_{1}$ polarizability is larger perpendicular to the Cl-O bond than parallel. 
Prior experiment and theory has shown that this is not the case in bulk 
\cite{Hyodo1989}. Our electronic structure calculations suggests that this is not the case for 
\ce{ClO4-} in an applied field similar in amplitude to what we would expect in the local interfacial environment.

%%%%
\subsubsection{Is ClO$_{4}^{-}$ orientation concentration dependent?}
\begin{figure}[htbp]
	\begin{center}
			\includegraphics[width=0.5\textwidth]{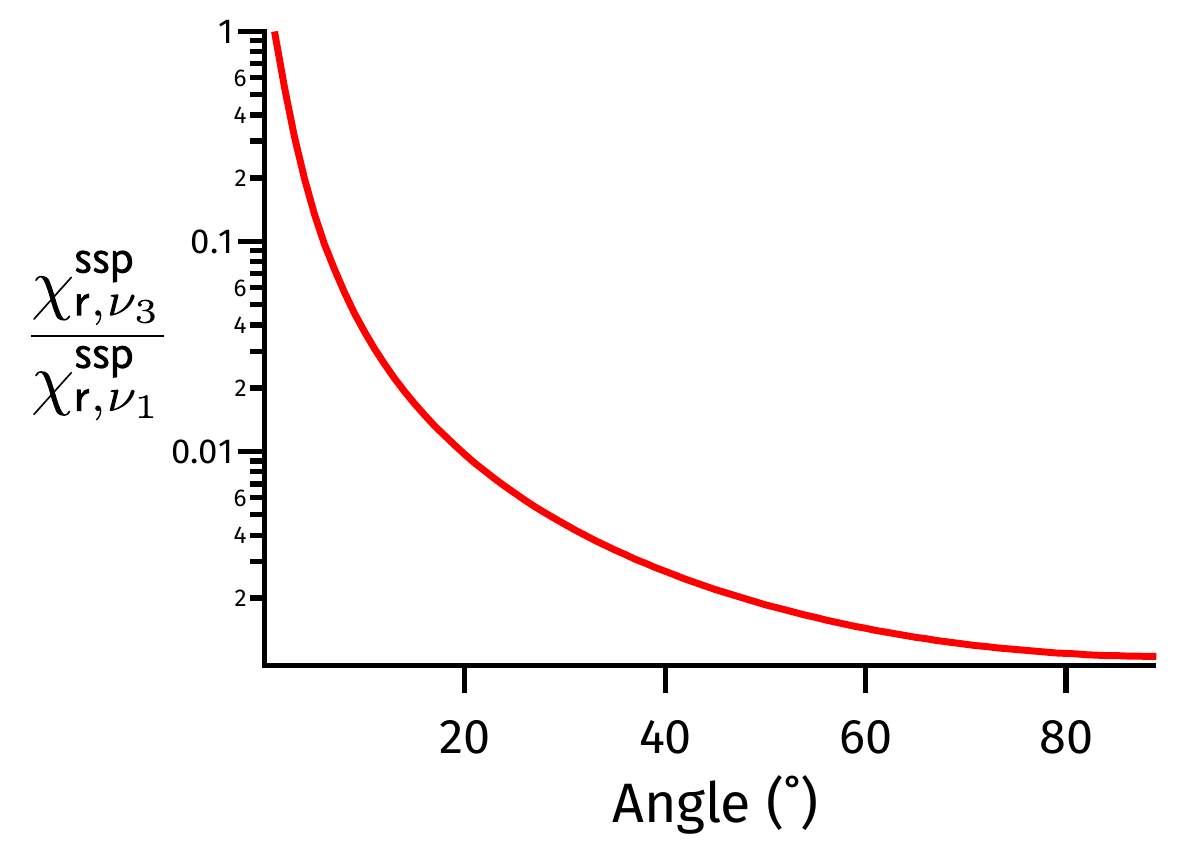}
		\caption{Calculated $\nicefrac{\chi^{\mathsf{ssp}}_{\mathsf{r},\nu_{3}}}
{\chi^{\mathsf{ssp}}_{\mathsf{r},\nu_{1}}}$ ratio as a function of \ce{ClO4-} 
orientation. Clearly even few degree changes in orientation should lead to large 
changes in ratio.}
		\label{f:nu31_calc}
	\end{center}
\end{figure}

\begin{figure}[htbp]
	\begin{center}
			\includegraphics[width=0.5\textwidth]{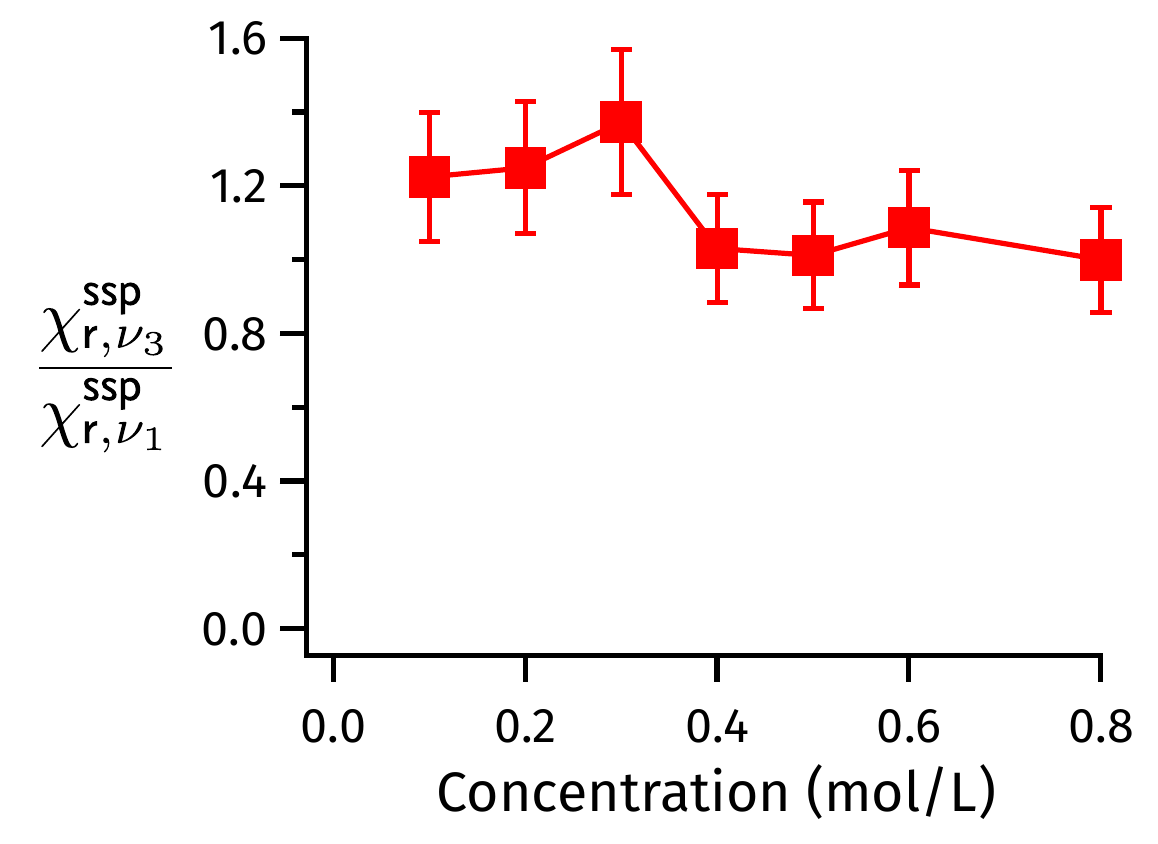}
		\caption{Ratios of $\nicefrac{\chi^{\mathsf{ssp}}_{\mathsf{r},\nu_{3}}}
{\chi^{\mathsf{ssp}}_{\mathsf{r},\nu_{1}}}$ extracted from experiment for spectra 
collected under the \textit{ssp} polarization condition. Clearly comparison of 
experiment with the calculated result shown in Figure \ref{f:nu31_calc} suggest 
that \ce{ClO4-} orientation is relatively insensitive to bulk \ce{HClO4} 
orientation.}
		\label{f:nu31_exp}
	\end{center}
\end{figure}

The transition dipole of the $\nu_{1}$ and $\nu_{3}$ modes of the \ce{ClO4-} anion 
are orthogonal. One consequence of this property is that changes in interfacial 
orientation of the \ce{ClO4-} anion will result in changes in relative intensities 
of the $\nu_{1}$ and $\nu_{3}$ modes as a function of bulk \ce{HClO4} orientation 
(for resonances appearing in spectra collected under a single polarization 
condition). The calculated dependence of the $\nicefrac{\chi^{\mathsf{ssp}}
_{\mathsf{r},\nu_{3}}}{\chi^{\mathsf{ssp}}_{\mathsf{r},\nu_{1}}}$ are shown in 
Figure \ref{f:nu31_calc} and the experimentally measured values, repeated from 
Figure 2 in the manuscript for ease of comparison, in Figure \ref{f:nu31_exp}. 
Comparison of the two figures makes clear that, if \ce{ClO4-} orientation changes 
as a function of bulk concentration of \ce{HClO4} (and thus presumably with increasing \emph{interfacial} \ce{ClO4-} concentration), this orientation change must be 
small.
%%%
%%%

%%%

\subsection{Computational Details}
\subsubsection{Methods}
We simulated the deformation of an isolated \ce{ClO4-} anion exposed to 
external electric dipole fields along the \textit{z}-axis and studied the resulting changes of the dipole moment and the related Raman depolarization ratio. Three popular density functional approximations, \textit{i.e.}\ PBEPBE\cite{per96}, PBE0\cite{per96b} and B3LYP\cite{bec93}, were employed together with a series of Gaussian-type basis sets, (aug-)cc-pV\emph{n}Z with \emph{n}=T, Q, 5. The calculations were performed using the GAMESS package\cite{sch93b}. As shown in Tables \ref{t:pbe_TZ}-\ref{t:p0_a5Z}, the three methods predict a very similar influence of the external electric dipole fields on the \ce{ClO4-} anion, with a slight method-dependent discrepancy in the calculated Cl-O bond length.

\subsubsection{Results}

%%%%%%%PBEPBE/cc-pVTZ
\begin{table}
	\begin{center}
	\footnotesize{
		\begin{tabular}{ccccccccc}
%			PBEPBE/cc-pVTZ &  &  &  &  &  &  &  &  \\ 
			E$_{z}$ (Debye) & $\phi_{z}$ (meV) & Dipole & $\rho$ & Cl-O$_{\text{z}}$ ({\AA}) & Cl-O$_{\text{other}}$ ({\AA}) & Angle & $\nicefrac{\text{Cl-O}_{\text{z}}}{\text{Cl-O}_{\text{other}}}$ & IR intens \\
			\toprule
			0 & 0 & 0.0001 & 0.0000 & 1.4916 & 1.4916 & 109.47 & 1.00 & 0.0001 \\ 
			0.0127 & 136.06 & -0.2316 & 0.0003 & 1.5021 & 1.4884 & 108.96 & 1.01 & 3.403 \\ 
			0.0254 & 272.11 & -0.4672 & 0.0018 & 1.5138 & 1.4852 & 108.46 & 1.02 & 17.68 \\ 
			0.0381 & 408.17 & -0.7089 & 0.0056 & 1.5275 & 1.4820 & 107.94 & 1.03 & 50.59 \\ 
			0.0508 & 544.23 & -0.9575 & 0.0124 & 1.5434 & 1.4788 & 107.44 & 1.04 & 103.1 \\
			\bottomrule
		\end{tabular}}
		\caption{Results employing the approach described above using the PBEPBE/cc-pVTZ model chemistry.}
		\label{t:pbe_TZ}
	\end{center}
\end{table}
%%%%%%%%

%%%%%%%PBEPBE/aug-cc-pVTZ
\begin{table}
	\begin{center}
	\footnotesize{
		\begin{tabular}{ccccccccc}
			E$_{z}$ (Debye) & $\phi_{z}$ (meV) & Dipole & $\rho$ & Cl-O$_{\text{z}}$ ({\AA}) & Cl-O$_{\text{other}}$ ({\AA}) & Angle & $\nicefrac{\text{Cl-O}_{\text{z}}}{\text{Cl-O}_{\text{other}}}$ & IR intens \\
			\toprule
			0 & 0 & 0.0001 & 0.0000 & 1.4954 & 1.4954 & 109.47 & 1.00 & 0.0002 \\ 
			0.0127 & 136.06 & -0.3041 & 0.0013 & 1.5074 & 1.4919 & 108.96 & 1.01 & 7.004 \\ 
			0.0254 & 272.11 & -0.6242 & 0.0082 & 1.5222 & 1.4882 & 108.40 & 1.02 & 43.15 \\ 
			0.0381 & 408.17 & -0.9588 & 0.0251 & 1.5397 & 1.4844 & 107.88 & 1.04 & 125.6 \\ 
			0.0508 & 544.23 & -1.302 & 0.0450 & 1.5603 & 1.4807 & 107.31 & 1.04 & 229.2 \\
			\bottomrule
		\end{tabular}}
		\caption{Results employing the approach described above using the PBEPBE/aug-cc-pVTZ model chemistry.}
		\label{t:pbe_aTZ}
	\end{center}
\end{table}
%%%%%%%%

%%%%%%%PBEPBE/aug-cc-pVQZ	
\begin{table}
	\begin{center}
	\footnotesize{
		\begin{tabular}{ccccccccc}
			E$_{z}$ (Debye) & $\phi_{z}$ (meV) & Dipole & $\rho$ & Cl-O$_{\text{z}}$ ({\AA}) & Cl-O$_{\text{other}}$ ({\AA}) & Angle & $\nicefrac{\text{Cl-O}_{\text{z}}}{\text{Cl-O}_{\text{other}}}$ & IR intens \\
			\toprule
			0 & 0 & 0.0001 & 0.0000 & 1.4831 & 1.4831 & 109.47 & 1.00 & 0.0002 \\ 
			0.0127 & 136.06 & -0.3015 & 0.0011 & 1.4946 & 1.4797 & 108.96 & 1.01 & 6.572 \\ 
			0.0254 & 272.11 & -0.6160 & 0.0077 & 1.5086 & 1.4758 & 108.46 & 1.02 & 42.05 \\ 
			0.0381 & 408.17 & -0.9482 & 0.0238 & 1.5248 & 1.4722 & 107.93 & 1.04 & 124.4 \\ 
			0.0508 & 544.23 & -1.312 & 0.0417 & 1.5452 & 1.4687 & 107.37 & 1.05 & 238.3 \\
			\bottomrule
		\end{tabular}}
		\caption{Results employing the approach described above using the PBEPBE/aug-cc-pVQZ model chemistry.}
		\label{t:pbe_aQZ}
	\end{center}
\end{table}
%%%%%%%%

%%%%%%%PBEPBE/aug-cc-pV5Z	
\begin{table}
	\begin{center}
	\footnotesize{
		\begin{tabular}{ccccccccc}
			E$_{z}$ (Debye) & $\phi_{z}$ (meV) & Dipole & $\rho$ & Cl-O$_{\text{z}}$ ({\AA}) & Cl-O$_{\text{other}}$ ({\AA}) & Angle & $\nicefrac{\text{Cl-O}_{\text{z}}}{\text{Cl-O}_{\text{other}}}$ & IR intens \\
			\toprule
			0 & 0 & 0.0001 & 0.0000 & 1.4731 & 1.4731 & 109.47 & 1.00 & 0.0002 \\ 
			0.0127 & 136.06 & -0.2982 & 0.0011 & 1.4845 & 1.4694 & 108.98 & 1.01 & 7.050 \\ 
			0.0254 & 272.11 & -0.6076 & 0.0074 & 1.4978 & 1.4660 & 108.48 & 1.02 & 41.95 \\ 
			0.0381 & 408.17 & -0.9369 & 0.0237 & 1.5137 & 1.4625 & 107.96 & 1.04 & 128.3 \\ 
			0.0508 & 544.23 & -1.298 & 0.0392 & 1.5328 & 1.4590 & 107.41 & 1.05 & 249.7 \\
			\bottomrule
		\end{tabular}}
		\caption{Results employing the approach described above using the PBEPBE/aug-cc-pV5Z model chemistry.}
		\label{t:pbe_a5Z}
	\end{center}
\end{table}
%%%%%%%%

%%%%%%%PB3LYP/aug-cc-pVTZ	
\begin{table}
	\begin{center}
	\footnotesize{
		\begin{tabular}{ccccccccc}
			E$_{z}$ (Debye) & $\phi_{z}$ (meV) & Dipole & $\rho$ & Cl-O$_{\text{z}}$ ({\AA}) & Cl-O$_{\text{other}}$ ({\AA}) & Angle & $\nicefrac{\text{Cl-O}_{\text{z}}}{\text{Cl-O}_{\text{other}}}$ & IR intens \\
			\toprule
			0 & 0 & 0.0001 & 0.0000 & 1.4789 & 1.4789 & 109.47 & 1.00 & 0.0001 \\ 
			0.0127 & 136.06 & -0.2878 & 0.0011 & 1.4909 & 1.4755 & 108.96 & 1.01 & 6.768 \\ 
			0.0254 & 272.11 & -0.5818 & 0.0066 & 1.5037 & 1.4720 & 108.45 & 1.02 & 37.19 \\ 
			0.0381 & 408.17 & -0.8911 & 0.0215 & 1.5195 & 1.4686 & 107.91 & 1.03 & 109.1 \\ 
			0.0508 & 544.23 & -1.219 & 0.0437 & 1.5378 & 1.4651 & 107.36 & 1.05 & 204.6 \\
			\bottomrule
		\end{tabular}}
		\caption{Results employing the approach described above using the B3LYP/aug-cc-pVTZ model chemistry.}
		\label{t:B_aTZ}
	\end{center}
\end{table}
%%%%%%%%

%%%%%%%PB3LYP/aug-cc-pV5Z	
\begin{table}
	\begin{center}
	\footnotesize{
		\begin{tabular}{ccccccccc}
			E$_{z}$ (Debye) & $\phi_{z}$ (meV) & Dipole & $\rho$ & Cl-O$_{\text{z}}$ ({\AA}) & Cl-O$_{\text{other}}$ ({\AA}) & Angle & $\nicefrac{\text{Cl-O}_{\text{z}}}{\text{Cl-O}_{\text{other}}}$ & IR intens \\
			\toprule
			0 & 0 & -0.0001 & 0.0000 & 1.4572 & 1.4572 & 109.47 & 1.00 & 0.0001 \\ 
			0.0127 & 136.06 & -0.2651 & 0.0009 & 1.4680 & 1.4538 & 108.99 & 1.01 & 6.892 \\ 
			0.0254 & 272.11 & -0.5390 & 0.0063 & 1.4807 & 1.4506 & 108.49 & 1.02 & 41.12 \\ 
			0.0381 & 408.17 & -0.8244 & 0.0205 & 1.4951 & 1.4473 & 107.99 & 1.03 & 119.6 \\ 
			0.0508 & 544.23 & -1.132 & 0.0415 & 1.5127 & 1.4437 & 107.46 & 1.05 & 230.6 \\
			\bottomrule
		\end{tabular}}
		\caption{Results employing the approach described above using the B3LYP/aug-cc-pV5Z model chemistry.}
		\label{t:B_a5Z}
	\end{center}
\end{table}
%%%%%%%%

%%%%%%%PBE0/aug-cc-pVTZ	
\begin{table}
	\begin{center}
	\footnotesize{
		\begin{tabular}{ccccccccc}
			E$_{z}$ (Debye) & $\phi_{z}$ (meV) & Dipole & $\rho$ & Cl-O$_{\text{z}}$ ({\AA}) & Cl-O$_{\text{other}}$ ({\AA}) & Angle & $\nicefrac{\text{Cl-O}_{\text{z}}}{\text{Cl-O}_{\text{other}}}$ & IR intens \\
			\toprule
			0 & 0 & 0.0001 & 0.0000 & 1.4620 & 1.4620 & 109.47 & 1.00 & 0.0141 \\ 
			0.0127 & 136.06 & -0.2737 & 0.0008 & 1.4722 & 1.4589 & 108.98 & 1.01 & 6.315 \\ 
			0.0254 & 272.11 & -0.5552 & 0.0054 & 1.4840 & 1.4558 & 108.49 & 1.02 &  36.47 \\ 
			0.0381 & 408.17 & -0.8460 & 0.0176 & 1.4973 & 1.4528 & 107.99 & 1.03 & 103.4 \\ 
			0.0508 & 544.23 & -1.150 & 0.0377 & 1.5124 & 1.4496 & 107.47 & 1.04 & 197.6 \\
			\bottomrule
		\end{tabular}}
		\caption{Results employing the approach described above using the PBE0/aug-cc-pVTZ model chemistry.}
		\label{t:p0_aTZ}
	\end{center}
\end{table}
%%%%%%%%

%%%%%%%PBE0/aug-cc-pV5Z	
\begin{table}
	\begin{center}
	\footnotesize{
		\begin{tabular}{ccccccccc}
			E$_{z}$ (Debye) & $\phi_{z}$ (meV) & Dipole & $\rho$ & Cl-O$_{\text{z}}$ ({\AA}) & Cl-O$_{\text{other}}$ ({\AA}) & Angle & $\nicefrac{\text{Cl-O}_{\text{z}}}{\text{Cl-O}_{\text{other}}}$ & IR intens \\
			\toprule
			0 & 0 & -0.0001 & 0.0000 & 1.4437 & 1.4437 & 109.47 & 1.00 & 0.0003 \\ 
			0.0127 & 136.06 & -0.2538 & 0.0007 & 1.4537 & 1.4408 & 109.01 & 1.01 & 5.870 \\ 
			0.0254 & 272.11 & -0.5151 & 0.0048 & 1.4651 & 1.4379 & 108.53 & 1.02 &  34.85 \\ 
			0.0381 & 408.17 & -0.7857 & 0.0166 & 1.4779 & 1.4349 & 108.05 & 1.03 & 104.6 \\ 
			0.0508 & 544.23 & -1.070 & 0.0359 & 1.4921 & 1.4316 & 107.56 & 1.04 & 207.6 \\
			\bottomrule
		\end{tabular}}
		\caption{Results employing the approach described above using the PBE0/aug-cc-pV5Z model chemistry.}
		\label{t:p0_a5Z}
	\end{center}
\end{table}
%%%%%%%%

\subsubsection{Solutions with Concentrations $>$ 1 M HClO$_{4}$}
\begin{figure}[htbp]
	\begin{center}
			\includegraphics[width=0.6\textwidth]{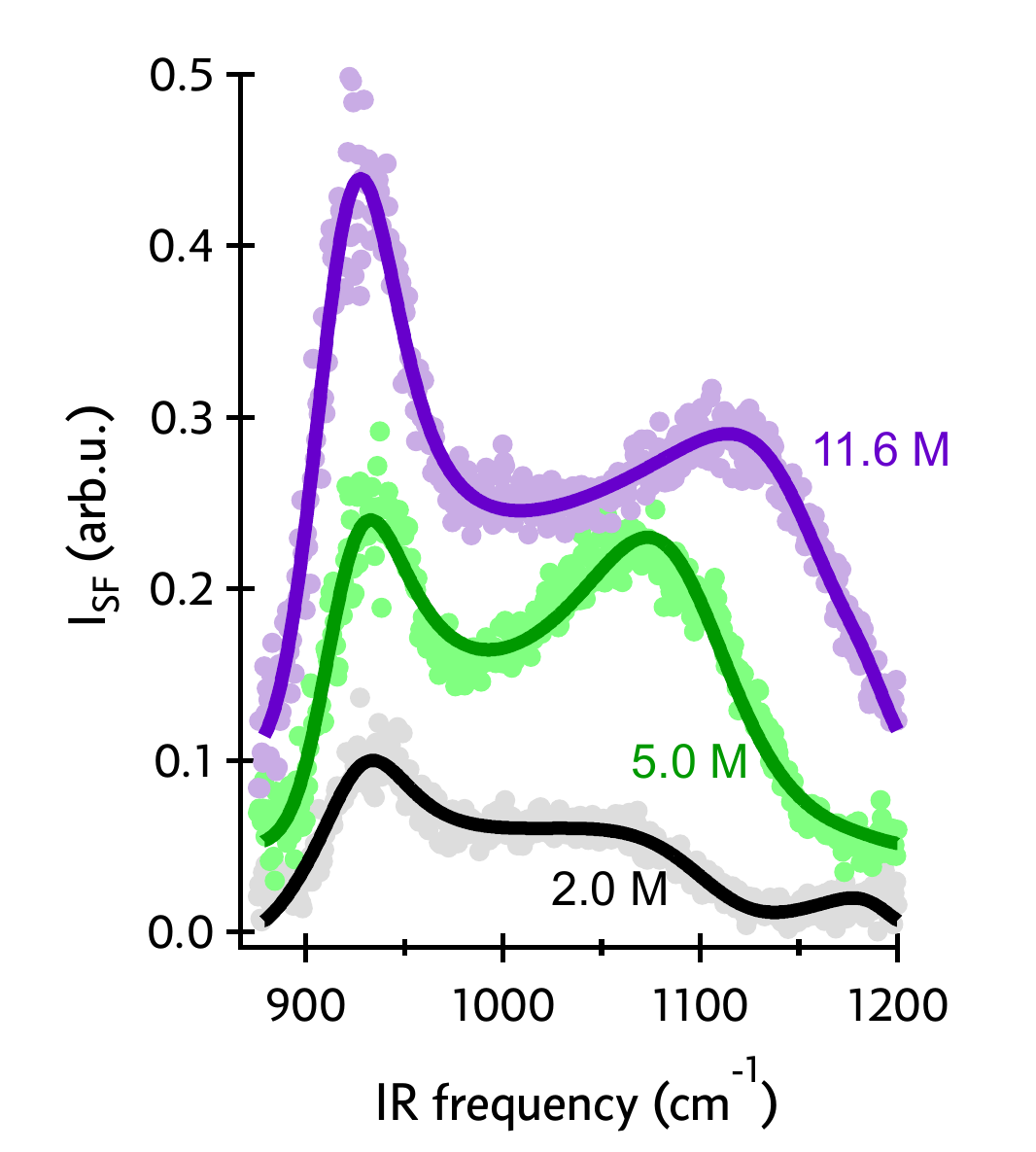}
		\caption{VSF spectra of bulk \ce{HClO4} solutions of 2, 5 and 11.6 M collected under the \textit{ssp} polarization condition. Dot are data, the solid lines are fits. In order to describe the data using the line shape model described above we found it necessary to introduce an additional resonance. Clearly, when comparing these data to those shown in Figure 1 in the manuscript, with increasing concentration the $\nu_{3}$ mode appears to split at sufficiently high concentration. Spectra are offset for clarity.}
		\label{f:highc}
	\end{center}
\end{figure}

I$_{\text{sf}}$ spectra of 2, 5 and 11.6 M solutions of \ce{HClO4} are shown in 
Figure \ref{f:highc}. These higher concentration spectra show a clear shift in the 
maximum of the spectral response associated with the $\nu_{3}$ mode and a gain in 
intensity between the $\nu_{1}$ and $\nu_{3}$. These trends are consistent with 
the splitting of $\nu_{3}$ expected under conditions in which $T_{d}$ symmetry is 
lifted. A similar gain in intensity at frequencies between the $\nu_{1}$ and $
\nu_{3}$ modes has been previously observed in Raman spectra of aqueous \ce{HClO4} 
solutions above 16 M and assigned to ion pairing or the appearance of molecular 
acid. Quantitative analysis suggests that these phenomena occur in 
\emph{interfacial} \ce{ClO4-} at concentrations more than 10$\times$ lower that in 
bulk.

\clearpage

\providecommand{\latin}[1]{#1}
\providecommand*\mcitethebibliography{\thebibliography}
\csname @ifundefined\endcsname{endmcitethebibliography}
  {\let\endmcitethebibliography\endthebibliography}{}

\end{document}